\documentclass[prx,a4paper,aps,twocolumn,superscriptaddress,10pt,nofootinbib]{revtex4-1}

\usepackage{amsmath,amsthm,amsfonts,graphicx,xcolor,times,xfrac,booktabs,mathtools,enumitem,xr,subfigure,amssymb,bbm,verbatim,appendix,placeins}
\usepackage[unicode=true,bookmarks=true,bookmarksnumbered=false,bookmarksopen=false,breaklinks=false,pdfborder={0 0 1}, backref=false,colorlinks=true]{hyperref}

\renewcommand*{\url}[1]{\href{#1}{#1}}

\newcommand{\inp}{\text{\normalfont \texttt{i}}}
\newcommand{\out}{\text{\normalfont \texttt{o}}}

\makeatletter
\theoremstyle{plain}
\newtheorem{thm}{\protect\theoremname}
\theoremstyle{plain}
\newtheorem{lem}[thm]{\protect\lemmaname}
\theoremstyle{plain}
\newtheorem{prop}[thm]{\protect\propositionname}
\theoremstyle{remark}
\newtheorem*{rem*}{\protect\remarkname}
\theoremstyle{plain}

\theoremstyle{plain}
\newtheorem{cor}[thm]{\protect\corollaryname}
\theoremstyle{definition}
\newtheorem{defn}[thm]{\protect\definitionname}
\theoremstyle{plain}
\newtheorem*{thm*}{\protect\theoremname}
\theoremstyle{plain}
\newtheorem*{lem*}{\protect\lemmaname}
 
\providecommand{\propositionname}{Proposition}
\providecommand{\theoremname}{Theorem}
\providecommand{\lemmaname}{Lemma}
\providecommand{\remarkname}{Remark}
\providecommand{\conjecturename}{Conjecture}
\providecommand{\definitionname}{Definition}
\providecommand{\corollaryname}{Corollary}
\allowdisplaybreaks

\newcommand{\trthm}{\normalfont \mathrm{tr}}

\def\bra#1{\langle{#1}\vert}
\def\ket#1{\vert{#1}\rangle}
\def\braket#1{\langle{#1}\rangle}

\def\BraVert{e.g.,roup\,\mid\,\bgroup}

\def\ketbra#1#2{\vert{#1}\rangle\!\langle{#2}\vert}

\def\tr#1{\mbox{tr}\left[{#1}\right]}
\newcommand{\ptr}[2]{\mbox{tr}_{#1}\left[ #2 \right]}

\DeclareMathOperator\sgn{sgn}

\def\uln#1{\underline{#1}}

\begin{document}


\title{Non-Markovian memory strength bounds quantum process recoverability}
\date{\today}

\author{Philip Taranto}
\email{philip.taranto@oeaw.ac.at}
\affiliation{Institute for Quantum Optics and Quantum Information --- IQOQI Vienna, Austrian Academy of Sciences, Boltzmanngasse 3, 1090 Vienna, Austria}
\affiliation{Atominstitut, Technische Universit{\"a}t Wien, 1020 Vienna, Austria}

\author{Felix A. Pollock}
\affiliation{School of Physics \& Astronomy, Monash University, Clayton, Victoria 3800, Australia}

\author{Kavan Modi}
\email{kavan.modi@monash.edu}
\affiliation{School of Physics \& Astronomy, Monash University, Clayton, Victoria 3800, Australia}


\begin{abstract}
Generic non-Markovian quantum processes have infinitely long memory, implying an exact description that grows exponentially in complexity with observation time. Here, we present a finite memory ansatz that approximates (or recovers) the true process with errors bounded by the strength of the non-Markovian memory. The introduced memory strength is an operational quantity and depends on the way the process is probed. Remarkably, the recovery error is bounded by the smallest memory strength over all possible probing methods. This allows for an unambiguous and efficient description of non-Markovian phenomena, enabling compression and recovery techniques pivotal to near-term technologies. We highlight the implications of our results by analyzing an exactly solvable model to show that memory truncation is possible even in a highly non-Markovian regime. 

\textbf{Keywords:} Open quantum systems; Stochastic processes; Non-Markovianity.
\end{abstract}

\maketitle


\section*{Introduction}

Our ability to manipulate quantum systems underpins potential advantages over classical technologies~\cite{NielsenChuang}. Their dynamics are often idealized as noiseless or, if unavoidable, noise is assumed to be uncorrelated. However, interactions with the environment generally perpetuate past information about the system to the future, thereby serving as a memory. Memory effects thus pervade physical evolutions, resulting in non-Markovian dynamics~\cite{StochProc}. The complexity of describing processes grows exponentially with memory length; therefore many simulation techniques invoke memory cutoffs~\cite{deVega2017}. Although several metrics have been proposed to quantify memory (and the consequences of neglecting it)~\cite{Li2018}, most do not consider the influence of interventions, overlooking the operational reality of sequentially probed dynamics. Indeed, the impact of memory depends on how a system is controlled~\cite{deVega2017,Li2018,Modi2011, Modi2012, Modi2012A, Milz2017, Milz2017KET,Pollock2018L,Pollock2018A,Taranto2019L,Taranto2019A,TarantoThesis}: generically, the system-environment state at any time is correlated, so an interrogation directly influences the system state and conditions the environment, both of which affect the future. Detected memory properties are thus naturally related to the interrogation method~\cite{TarantoThesis}. This operational perspective has important consequences for dynamical decoupling~\cite{Viola1999, Arenz2015, Arenz2017}, erasure or transmission of information~\cite{Kretschmann2005, Caruso2014}, correlated error correction and characterization~\cite{Morris2019,romero2021}, and (operational) quantum thermodynamics~\cite{Romero2018, Strasberg2018, Romero2019, Strasberg2019-1, Strasberg2019-2, Romero2020}.

\begin{figure}[t]
\centering
\vspace{0.5em}
\includegraphics[width=0.95\linewidth]{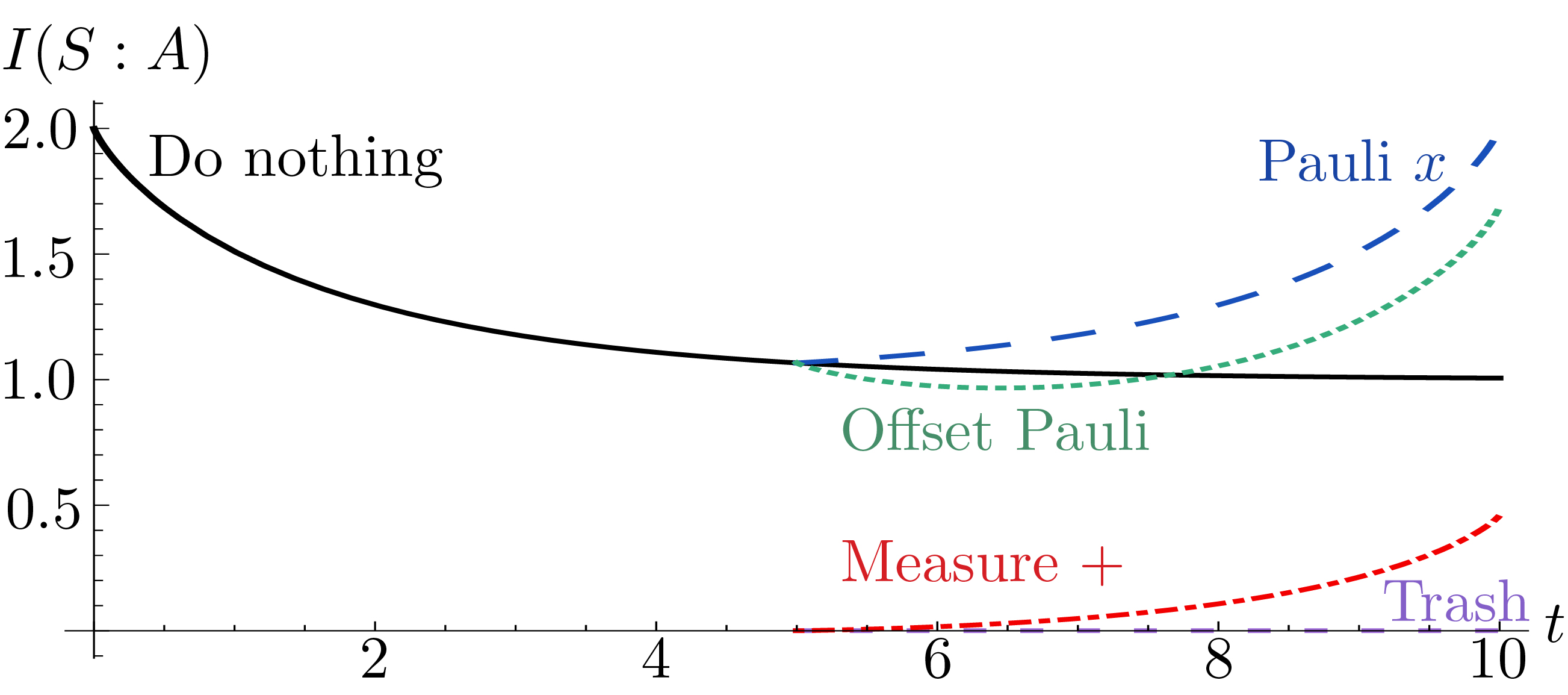}
\caption{\emph{Memory in shallow pocket dynamics.} The mutual information $I(S:A)$ between a system and ancilla initially in a Bell pair decays exponentially as the system undergoes shallow pocket evolution (black). However, this is not the case if an intervention is applied at $t_1$ ($=5$ above). We depict $\sigma_x$ (blue); an offset rotation $\sqrt{0.95}\sigma_x+\sqrt{0.05}\sigma_z$ (green); measurement of $+$ in the $x$-basis (red); and trash-and-reprepare (purple). } \label{fig:shallowpocket}
\end{figure}

This can be illustrated with the shallow pocket model~\cite{Lindblad1980, Accardi1982, Arenz2015, Arenz2017}, comprising a qubit coupled to a continuous degree of freedom (see Fig.~\ref{fig:shallowpocket} and Appendix~\ref{app:shallowpocket}). The joint dynamics induces pure-dephasing Lindblad evolution for the qubit, with exponentially-decaying coherences. Non-classical correlations between any preparation and measurement similarly vanish, so the reduced dynamics forgets the initial state. However, the evolution following a $\sigma_x$ unitary reverts the system to its original state; in this sense, the process displays infinitely-long memory. This example highlights that, although certain temporal correlations of the unperturbed system may decay rapidly, they do not account for the whole story; more generally, there exist detectable correlations between the history and future \emph{processes}. 

Disentangling the non-Markovian memory, carried by the environment, from the temporal correlations that result from probing the system is a challenge that has only recently been resolved through an operational framework for describing quantum stochastic processes~\cite{Pollock2018L,Pollock2018A}. Subsequently, a notion of memory length or quantum Markov order was introduced, which reduces the complexity of describing processes with short-term memory by only retaining the minimal number of recent timesteps relevant to the dynamics~\cite{Taranto2019L,Taranto2019A}. Higher-order Markov models capture all memory effects below the Markov order and are therefore more accurate than Markovian approximations. However, a missing element from this theory so far has been a method for quantifying the strength of non-Markovian memory and developing efficient approximations accordingly: truncating weak temporal-correlations will yield a more efficient description for a process without compromising the accuracy. 

In this article, we construct an operational notion of memory strength for quantum processes. That is, we quantify the correlations between history and future processes with respect to an intermediate (multi-time) probing schema (see Fig.~\ref{fig:processtensor}). Our main result links said memory strength with process recoverability: with respect to any interrogation sequence, one can approximate the process by discarding future-history correlations. If the memory strength is small for some instrument (family), then the approximate process accurately predicts expectation values for related observables, namely those in the linear span of the original instrument elements. This connection is akin to that between conditional mutual information and fidelity of recovery for quantum states~\cite{Fawzi2015, Sutter2016}, and involves a generalization of the measured relative entropy~\cite{Piani2009} to quantum stochastic processes. A corollary follows for the `do nothing' instrument---which many approximations assume---which bounds the accuracy of predicting future states. Moreover, the memory strength for an informationally-complete instrument bounds the distinguishability between the actual and recovered process for \emph{any} experimental protocol. Lastly, we demonstrate our results via a solvable non-Markovian model, highlighting the complex memory structures amenable to our framework. 

\section*{Background}

\subsection{Classical Stochastic Processes}

\begin{figure}[t]
\centering
\includegraphics[width=0.95\linewidth]{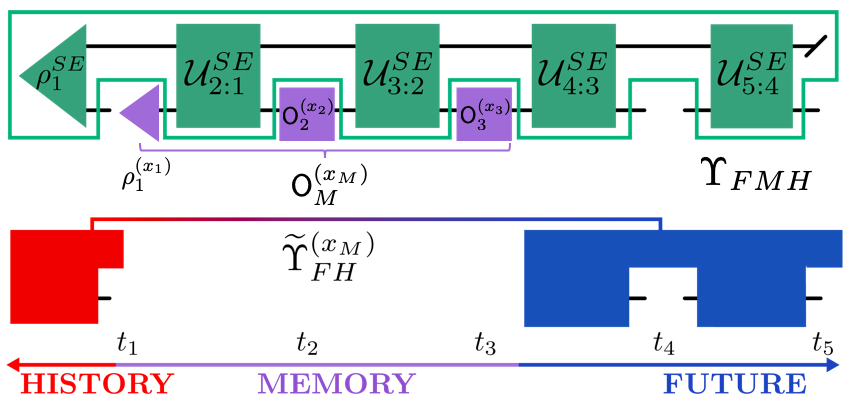}
\caption{\emph{Quantum stochastic process.} An open quantum process with an initial system-environment state and unitary evolutions (green) can be represented as a process tensor $\Upsilon_{FMH}$ (outline). For each event of a memory instrument sequence, $\mathcal{J}_M = \{ \mathsf{O}_M^{(x_M)}\}$ (purple), a conditional future-history process $\widetilde{\Upsilon}_{FH}^{(x_M)}$ (blue, red) results; future-history correlations evidence memory effects across $\ell := |M|$. \label{fig:processtensor}}
\end{figure}

We begin by reviewing the pertinent ingredients from the theory of classical stochastic processes before turning our attention to quantum stochastic processes. A classical stochastic process over a discrete set of times $\mathcal{T} := \{ t_1, \hdots, t_n\}$ is described by an $n$-point joint probability distribution $\mathbbm{P}(x_n,\hdots,x_1)$. A process has finite-length memory whenever the probability of each event $x_k$ at time $t_k \in \mathcal{T}$ only conditionally depends upon the past $\ell$ events:
\begin{gather} \label{eq:clst}
    \mathbbm{P}(x_k|x_{k-1},\hdots,x_1)=\mathbbm{P}(x_k|x_{k-1},\hdots,x_{k-\ell}).
\end{gather}
Here $\ell$, the minimum number for which Eq.~\eqref{eq:clst} holds, denotes the Markov order; a Markovian process has $\ell\leq 1$. Markov order captures the complexity of characterizing a process, which grows exponentially in $\ell$. Although $\ell$ may be large for many processes, their memory can be truncated, permitting efficient approximation. Grouping the times into three segments: the history $H = \{ t_{1}, \hdots, t_{k-\ell-1}  \}$, memory $M = \{  t_{k-\ell}, \hdots, t_{k-1} \}$ and future $F = \{ t_k , \hdots, t_n\}$ (see Fig.~\ref{fig:processtensor}), Markov order $\ell$ implies the conditional factorization
\begin{gather}\label{eq:cmarkovcondindep}
    \mathbbm{P}_{FH}(x_F,x_H|x_M) = \mathbbm{P}_{F}(x_F|x_M) \mathbbm{P}_{H}(x_H|x_M),
\end{gather}
i.e., the future and history are conditionally independent given memory events.\footnote{Here, the conditional probability distribution over the history is post-selected on memory events, which is similarly the case in the generalization to the quantum setting.} This is equivalently expressed by the vanishing classical conditional mutual information \textbf{(CMI)}, 
\begin{gather}\label{eq:clcmi}
I(F:H|M) := \mathsf{h}_{FM} + \mathsf{h}_{MH} - \mathsf{h}_{FMH} - \mathsf{h}_{M},    
\end{gather}
where $\mathsf{h}_{X} := -\sum_x \mathbb{P}_X(x) \log[\mathbb{P}_X(x)]$. The CMI is interpreted as the memory strength of the process. For processes with Markov order $\ell$, $I(F:H|M) = 0$ by Eq.~\eqref{eq:cmarkovcondindep}; however, in general $\mathbbm{P}_{FMH}$ does not conditionally factorize, and the CMI quantifies the correlations between $F$ and $H$, given $M$.

The significance of approximately-finite Markov order is best encapsulated through the \emph{recovery map}, $\mathcal{R}_{M \to FM}$, that acts only on $M$ to approximate the correct future statistics: $\mathbbm{P}_{FMH}(x_F, x_M, x_H) \simeq \mathcal{R}_{M \to FM} [\mathbbm{P}_{MH}(x_M, x_H)]$, with equality holding for processes with Markov order $\leq \ell$. Intuitively, the recovery map discards conditional future-history correlations and uses the r.h.s. of Eq.~\eqref{eq:cmarkovcondindep} as an approximate description. Importantly, the recovery map approximates the process with an error that is bounded by the CMI~\cite{Fawzi2015, Sutter2016}, with complexity reduced to the approximate Markov order. Thus, whenever the memory is weak, the recovery map provides an accurate and efficient approximation.

\subsection{Quantum Stochastic Processes}

We now move to the realm of quantum stochastic processes. To quantify the strength of memory in quantum processes, we begin by introducing the process tensor framework, detailed in Methods, which generalizes Eq.~\eqref{eq:clst} to the quantum domain. Consider a joint system-environment $SE$, with (correlated) initial state $\rho^{SE}_1$. The system is interrogated at $t_1$ and a quantum \emph{event} $x_1$ is observed, with an associated completely-positive \textbf{(CP)} transformation $\mathcal{O}_1^{(x_1)}$ on $S$. The events are described by an \emph{instrument} $\mathcal{J}_1 =\{\mathcal{O}_1^{(x_1)}\}$, which is trace-preserving \textbf{(TP)}, i.e., $\mathcal{O}_1^{\mathcal{J}_1} := \sum_{x_1} \mathcal{O}_1^{(x_1)}$ is CPTP~\cite{Lindblad1979}. Following this intervention, $SE$ evolves unitarily for time $t_2 - t_1$ according to the superoperator $\mathcal{U}^{SE}_{2:1}$. Then, $S$ is probed at $t_2$, and so on, until $t_n$. The probability to observe $x_1, \hdots, x_n =: x_{n:1}$ using instruments $\mathcal{J}_1, \hdots, \mathcal{J}_n =: \mathcal{J}_{n:1}$ is
\begin{align}
    \mathbbm{P}(x_{n:1}|\mathcal{J}_{n:1})\!=
    &\tr{\mathcal{O}_n^{(x_n)} \mathcal{U}_{n:n-1}^{SE} \hdots \mathcal{U}_{2:1}^{SE} \mathcal{O}_1^{(x_1)}\! \rho^{SE}_1}.
\end{align}
\noindent As per Fig.~\ref{fig:processtensor}, abstracting everything outside of experimental control defines the process itself and yields a multi-linear map from instruments to probability distributions called the \emph{process tensor}~\cite{Pollock2018L, Pollock2018A}. On the other hand, the instruments on $S$ are collected to define the generalized instrument $\mathcal{J}_{n:1} = \{ \mathcal{O}_{n:1}^{(x_{n:1})} \}$. A generalized instrument can be temporally correlated, although we focus on uncorrelated instrument sequences for simplicity. Intuitively, the process tensor encapsulates the uncontrollable effect of the environment, i.e., the process per se, whereas the generalized instrument represents the controllable influence. Any open quantum dynamics probed at a (causally-ordered) number of times can be described by a process tensor, and any probing sequence by a generalized instrument~\cite{Chiribella2009, Milz2017}. (These objects are commonly known as quantum combs and have appeared elsewhere~\cite{Lindblad1979, Accardi1982, Kretschmann2005, Chiribella2008, Chiribella2008-2, Chiribella2009, Hardy2012, Oreshkov2012, Costa2016, Hardy2016,coecke_kissinger_2017}.)

Both the process tensor and the instrument elements are higher-order quantum maps~\cite{Chiribella2008-2, Chiribella2009,Milz2017,Bisio2019} that can respectively be represented as quantum states $\Upsilon_{n:1}$ and $\{\mathsf{O}_{n:1}^{(x_{n:1})}\}$ via the Choi-Jamio{\l}kowski isomorphism. The joint probability of realizing any sequence of events is given by the generalized Born-rule~\cite{ShrapnelCosta2017} (see Methods):
\begin{align}\label{eq:processtensor}
    \mathbbm{P}_{n:1}(x_{n:1}|\mathcal{J}_{n:1}) = \tr{ \mathsf{O}_{n:1}^{(x_{n:1}) \textup{T}} \Upsilon_{n:1}} 
    =: \langle
    \mathsf{O}_{n:1}^{(x_{n:1})}\rangle_{\Upsilon_{n:1}}.
\end{align} 
The process tensor encodes all probabilities for any choice of instruments and thus characterizes the process. Grouping the times as before, the quantum generalization of the l.h.s. of Eq.~\eqref{eq:cmarkovcondindep} is obtained by projecting the memory of $\Upsilon_{FMH}$ onto the conditioning element of $\mathcal{J}_{M} = \{ \mathsf{O}^{(x_M)}_M \}$. The result is the \emph{conditional future-history process} (see Fig.~\ref{fig:processtensor})
\begin{gather}
\label{eq:conditionalfuturehistory}
	\widetilde{\Upsilon}_{FH}^{(x_M)} = \ptr{M}{\mathsf{O}^{(x_M) \textup{T}}_M \Upsilon_{FMH}}.
\end{gather}

The tilde in Eq.~\eqref{eq:conditionalfuturehistory} signifies that the conditional objects are not necessarily proper process tensors, since realizing memory events post-selects the history~\cite{Chiribella2008, Milz2018}; nonetheless, summing these yields a proper process tensor $\Upsilon_{FH}^{\mathcal{J}_M} := \sum_{x_M} \widetilde{\Upsilon}_{FH}^{(x_M)}$. Mirroring the classical setting, if the conditional processes are uncorrelated, i.e., $\widetilde{\Upsilon}_{FH}^{(x_M)} = \Upsilon_{F}^{(x_M)} \otimes \widetilde{\Upsilon}_{H}^{(x_M)}$, the process has Markov order $\ell$ with respect to $\mathcal{J}_M$~\cite{Taranto2019L} [see r.h.s. of Eq.~\eqref{eq:cmarkovcondindep}].\footnote{Note that the causality constraint on the process tensor ensures that $\Upsilon_F^{(x_M)}$ is causally-ordered for each event.} This operational notion of memory strength means that by applying $\mathcal{J}_M$, no future-history correlations are possible for any history and future instruments, i.e., no memory lasting longer than $\ell$ is detectable. In general, the conditional processes are correlated. However, long memory does not imply strong memory; we now show how weak correlations can be truncated for accurate and efficient approximation.

\section*{Results}
\setcounter{subsection}{0}

The first ingredient to developing finite Markov order approximations is to quantify the memory strength over a block of length $\ell$, which we provide in Eq.~\eqref{eq:instrumentstrength}. We then construct an approximate process which neglects long-term memory. We first use the knowledge from the interrogation to (partially) tomographically reconstruct the process on the memory block, ensuring that the approximate process acts correctly on all instruments lying in the span of the original one. This reconstruction typically exhibits future-history correlations, which are subsequently discarded to yield the approximate process of Eq.~\eqref{eq:restricted}. We then bound the error of expectation values computed with the approximate process in terms of the memory strength, thereby endowing it with operational meaning. By iterating our procedure over translations of the memory block, significant savings in the complexity of description are possible~\cite{White2021,White2021Diagnosing}.

\subsection{Memory Strength and Recoverability}

Given a particular realization $x_M$ of instrument $\mathcal{J}_M$, the future and history can be more or less correlated. Such correlations can be detected by applying any choice of instruments $\mathcal{J}_F, \mathcal{J}_H$ on the conditional future-history process. Taking the supremum over such instruments provides an operational definition of memory strength, quantifying the largest detectable conditional future-history correlations. We thus define
\begin{align} \label{eq:instrumentstrength}
    \Theta(\mathcal{J}_M) \! := \!\!\! \sup_{\mathcal{J}_F,\mathcal{J}_H \in \mathbbm{J}} \sum_{x_M} p(x_M|\mathcal{J}_M)I_{\mathbbm{J}}(F:H|x_M)_\Upsilon ,
\end{align} 
where $p(x_M|\mathcal{J}_M):=\tr{\widetilde{\Upsilon}_{FH}^{(x_M)}}$ is the probability of observing $x_M$ and the supremum is taken over uncorrelated instruments $\mathcal{J}_Y := \{ \mathsf{O}_Y^{(x_Y)}\}$ for $Y \in \{ F, H\}$ belonging to a set $\mathbbm{J}$. We have defined the \emph{measured conditional mutual information}
\begin{align}\label{eq:measuredmutualinfo}
    &I_{\mathbbm{J}}(F:H|x_M)_\Gamma := I(F:H)_{\mathbbm{P}_{\Gamma}(x_F,x_H|x_M)}, \quad \mbox{where}\\
    &\mathbbm{P}_{\Gamma}(x_F,x_H|x_M) := \tr{(\mathsf{O}_F^{(x_F)} \otimes \mathsf{O}_H^{(x_H)})^\textup{T} \widetilde{\Gamma}_{FH}^{(x_M)} }
\end{align}
is the conditional probability distribution for any process $\Gamma$ and (independent) future-history instruments conditioned on a fixed memory event. Intuitively, the memory strength in Eq.~\eqref{eq:instrumentstrength} captures the largest detectable future-history correlation, conditioned on the outcomes recorded on the memory block, aggregated to the level of $\mathcal{J}_M$ by averaging over $x_M$.

Unlike classical memory, quantum memory effects depend upon the choice of probing scheme~\cite{Taranto2019L,Taranto2019A}, suggesting that a universal quantum recovery map may not exist. However, we now construct a quantum process recovery map that efficiently builds up longer processes from shorter ones when the process is stationary. We begin with an ansatz process with finite quantum Markov order with respect to $\mathcal{J}_M = \{\textsf{O}^{(x_M)}\}$:
\begin{gather} \label{eq:restricted}
    \uln{\Lambda}^{\mathcal{J}_M}_{FMH} := \sum_{x_M}\Upsilon_F^{(x_M)}\otimes \mathsf{D}^{(x_M)}_M\otimes\widetilde{\Upsilon}_{H}^{(x_M)},
\end{gather}
where the $\mathsf{D}^{(x_M)}_M$ are dual operators satisfying $\mbox{tr}[\mathsf{D}^{(x_M)\textup{T}}_M\textsf{O}^{(x_M')}_M] = \delta_{x_M x'_M}$~\cite{Modi2012A, Milz2017}. This is the tomographic representation of the process via linear inversion of the instrument outcomes, with conditional $FH$ correlations discarded. The recovered process can exhibit correlations between $H$ and $M$, $M$ and $F$ (as well as within each block), but \emph{not} between $H$ and $F$. By construction, $\uln{\Lambda}^{\mathcal{J}_M}_{FMH}$ is positive on its domain, which is the span of $\mathcal{J}_M$. When $\mathcal{J}_M$ is not informationally-complete, i.e., does not span the full space, then $\uln{\Lambda}^{\mathcal{J}_M}_{FMH}$ only approximates the original process in said subspace, which we denote by the underline. Such processes are called restricted process tensors~\cite{Milz2018A} and are commonly encountered in experiments~\cite{White2020,Taranto2020Exp,White2021,White2021Diagnosing}. Nevertheless, its action on its domain is guaranteed to reproduce the correct statistics for any multi-time observable of the form $C = \sum_{x} c_x^\mathcal{J} \mathsf{O}^{(x)}$ with $\mathsf{O}^{(x)} = \sum_{x_M} \! \mathsf{E}^{(x,x_M)}_{FH} \!\otimes \mathsf{O}^{(x_M)}_{M}$, with arbitrary $\mathsf{E}^{(x,x_M)}_{FH}$. This is a consequence of linearity, as the observable form ensures a linear decomposition in terms of $\mathsf{O}_M^{(x_M)}$, upon which the recovered process acts correctly due to $\mbox{tr}[\mathsf{D}^{(x_M)\textup{T}}_M\textsf{O}^{(x_M')}_M] = \delta_{x_M x'_M}$.

The above ansatz is the process analogue of a quantum Markov chain state~\cite{Fawzi2015, Sutter2016}, which is widely studied in the context of Petz's recovery map~\cite{Petz1986,Petz2003}. Like a quantum Markov chain state, the process above is generally not separable, as $F$ and $H$ can share entanglement with $M$, as per the example in Appendix~\ref{app:qmoexample}. However, since the dual elements here can be non-positive and cannot necessarily be decomposed into orthogonal parts, care must be taken in defining the recovery map $\mathcal{R}^{\mathcal{J}_M}_{M\rightarrow FM}: \uln{\Lambda}_{MH}^{\mathcal{J}_M} \to \uln{\Lambda}_{FMH}^{\mathcal{J}_M}$ (see Appendix~\ref{app:recovery}). The concept of the quantum recovery map is analogous to the classical case [see below Eq.~\eqref{eq:clcmi}]. The key advantage of the quantum process recovery map is that its repeated action on the ansatz propagates the process arbitrarily far into the future with fixed $\ell$-dependent complexity. If the process $\Upsilon_{FMH}$ has weak memory $\Theta(\mathcal{J}_M)$, the expectation value of any valid observable calculated from the recovered process $\uln{\Lambda}_{FMH}^{\mathcal{J}_M}$ accurately approximates that of the original:

\begin{thm}\label{thm:recovery}
For any multi-time observable $C$ with support on $M$ within the span of the elements of $\mathcal{J}_M$,
\begin{gather}
    \left|\langle C\rangle_{\Upsilon_{F\!M\!H}} - \langle C\rangle_{ \uln{\Lambda}^{\mathcal{J}_M}_{F\!M\!H}}\right| \leq |\mathbf{C}| \sqrt{2 \Theta(\mathcal{J}_M)},
    \label{eq:theorem}
\end{gather}
with $|\mathbf{C}| := \inf_{\mathcal{J}}\! \sqrt{\sum_{x}|c_x^\mathcal{J}|^2}$.
\end{thm}
\noindent This and the following statements are proven in the Methods section. Thm.~\ref{thm:recovery} is fully general inasmuch as it holds without any assumptions on the dynamics or instruments employed. The r.h.s. involves a supremum over instruments on the future and history; as the memory strength takes the form of a generalized divergence, recent numerical techniques can be used for its estimation~\cite{FawziOpt2018,FawziSDP2019,Fang2021}. In Appendix~\ref{app:proxies}, we provide an easier-to-compute (and looser) bound based on the relative entropy between the original and recovered process---which foregoes the requirement for optimization---by adapting results from Refs.~\cite{Piani2009, Fawzi2015, Petz2003, Paulsen2002} to first bound a generalized measured relative entropy and then the left-hand side of Eq.~\eqref{eq:theorem} via Pinsker's inequality. We also prove another bound, which is tighter in some cases, by restricting to unbiased instruments satisfying $\trthm_{M}\left[ \mathsf{O}^{\mathcal{J}}_{FMH}\right] \propto \mathbbm{1}_{FH}$. Such instruments have the unconditional action of a completely depolarising channel, e.g., a randomly sampled Clifford gate. Deriving tighter bounds under various physically-motivated (e.g., thermodynamic) assumptions remains an open problem.

While Thm.~\ref{thm:recovery} applies to multi-time observables, often one only requires the time-evolved density operator; a corollary bounds its prediction error:
\begin{cor}\label{cor:densityop}
Let $\rho_j^{(x_M)}$ be the true density operator at any time $t_j \in F$ following outcome $x_M$ of $\mathcal{J}_M$ applied to the memory, and ${\rho'}^{(x_M)}_j$ be the approximated one. Then:
\begin{gather}
    \left\|\rho_j^{(x_M)} - {\rho'}^{(x_M)}_j\right\|_1  \leq \sqrt{2\Theta(\mathcal{J}_M)}\; \forall \;t_j\in F.
\end{gather}
\end{cor}
\noindent The future states result from applying identity maps at all history and future times except $t_j$ and the instrument $\mathcal{J}_M$ to the memory block to the true and recovered process.  See Ref.~\cite{White2021} for a detailed analysis of using finite Markov order approximations to accurately prepare future states.

Whenever $\mathcal{J}_M$ spans the full space, i.e., is informationally complete, then any multi-time expectation value can be accurately approximated. In this case the distinguishability, by any means, is bounded by the memory strength:
\begin{thm}\label{thm:ic}
For informationally-complete $\mathcal{J}_M$, the recovered process $\Lambda^{\mathcal{J}_M}_{FMH}$ gives sensible predictions for any instrument on $M$ and 
\begin{gather} \label{eq:diamonddistance}
    \left\|\Upsilon_{FMH} - \Lambda_{FMH}^{\mathcal{J}_M} \right\|_\diamond\leq \sqrt{2 \Theta(\mathcal{J}_M)},
\end{gather}
where $\|X\|_\diamond := \sup_{\mathcal{J} = \{\mathsf{O}^{(x)}\}}\|\sum_x\trthm{[\mathsf{O}^{(x)} X\otimes \mathbbm{1}}]\ketbra{x}{x}\,\|_1$ generalizes the diamond norm\normalfont{~\cite{Gilchrist2005}} \emph{to quantum processes.} 
\end{thm}

\noindent The memory strength thus provides an operationally-clear measure: if there exists some informationally-complete instrument for which $\Theta(\mathcal{J}_M)$ is small, then Thm.~\ref{thm:ic} states that one can closely approximate the process for all instruments, even those for which the memory strength is large. If, additionally, $\Upsilon_{FH}^{(x_M)} \approx \Upsilon_{FH}^{(x'_M)} \ \forall \ x_M, x'_M$ in the informationally-complete instrument, then the process has small memory strength for all instruments; such processes resemble similar properties to approximately finite-memory classical stochastic processes (where there is only one instrument). 


\subsection{Case Study}

Consider a qubit $S$ coupled to another qubit $E$, which is cooled by an external bath. The joint evolution follows
\begin{align}
	\frac{\partial \rho^{SE}_t}{\partial t} = -i \xi [\sigma_x^S \otimes \sigma_x^E , \rho^{SE}_t] + \kappa \mathcal{L}[\sigma_{-}^E]( \rho^{SE}_t),
\end{align}
where the dissipator acts on $E$: $\mathcal{L}[\sigma_{-}^E]( \rho^{SE}_t) := \sigma_{-}^{E} \rho_t^{SE} \sigma_+^{E} - \tfrac{1}{2} \{  \sigma_+^{E}  \sigma_-^{E}, \rho^{SE}_t \}$, with $\sigma_{\pm}^{E} := \sigma_x^E \pm i \sigma_y^E$. In Ref.~\cite{Pang2017}, it was shown that for $\kappa^2 \geq 64 \xi^2$, the process is CP-divisible, which is a common proxy for quantum Markovianity~\cite{Rivas2014, Breuer2016}; however, CP-divisibility only implies an absence of some kinds of memory~\cite{Milz2019,Hsieh2019}. Non-Markovianity `measures' built upon two-time considerations---many of which are contradictory~\cite{Li2018}---overlook multi-time effects. Indeed, this model contains higher-order correlations; by constructing the process tensor, we quantify the (non-vanishing) non-Markovianity for all $(\xi,\kappa) \in [0,2] \times [0,10]$ (see Appendix~\ref{app:casestudy}). We then examine the memory strength in various regimes by constructing three 6-time process tensors $\Upsilon_{6:1}(\xi,\kappa)$: one {CP}-divisible, one strongly non-Markovian, and one intermediate, and let $M$ range from $t_2$ to $t_5$. We consider: i) the identity map, which captures the \emph{natural memory strength}; ii) the ``causal break'' instrument, where the system is measured and independently reprepared in an informationally-complete manner, breaking information flow through the system; and iii) the completely-noisy instrument, which replaces the state with white noise, quantifying \emph{noise-resistant memory}~\cite{TarantoThesis}. 

\begin{figure}[t]
\centering
\includegraphics[width=0.95\linewidth]{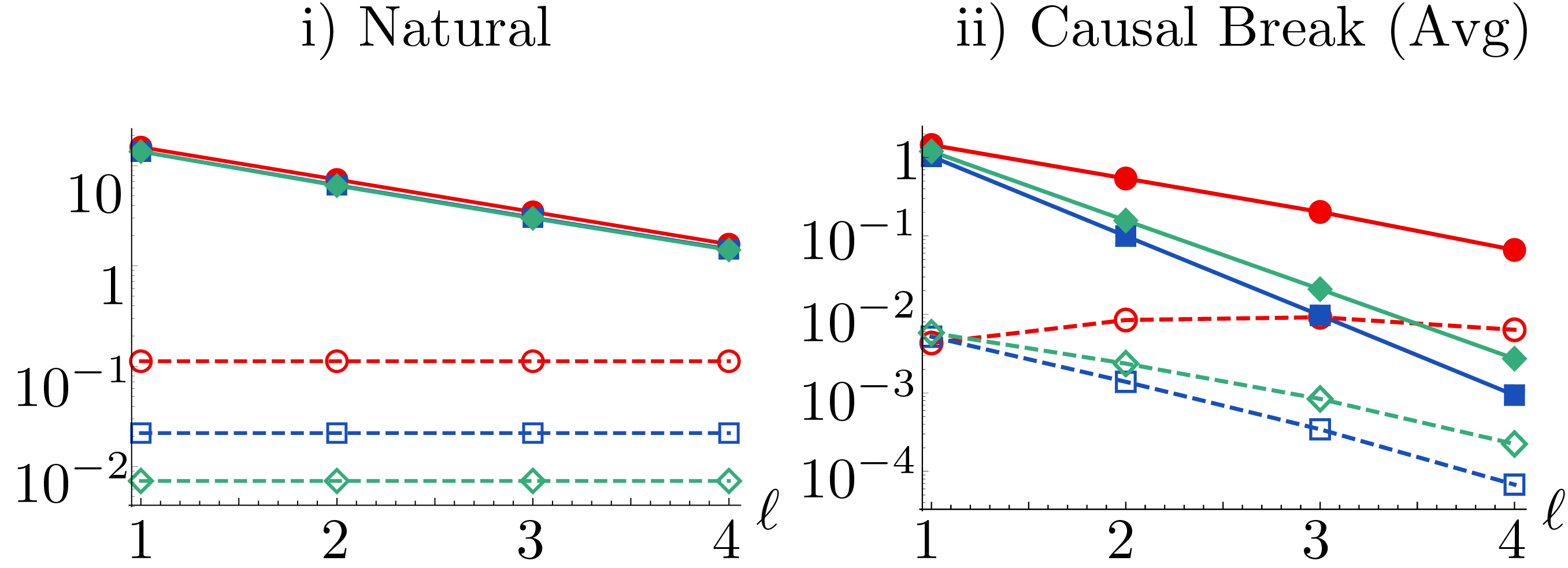}
\caption{\emph{Case study.} We plot $|\langle C\rangle_{\Upsilon_{F\!M\!H}} - \langle C\rangle_{ \uln{\Lambda}^{\mathcal{J}_M}_{F\!M\!H}}|$ (hollow, dashed) and the proxy memory strength based on the relative entropy between the Choi states of the true and recovered processes (see Appendix~\ref{app:proxies}) (solid) for i) the identity map and ii) a causal break. We construct a 6-step process tensor in the strongly non-Markovian (red, circles), CP-divisible (blue, squares) and intermediate (green, diamonds) regimes, and consider the memory strength over $\ell$ applications of said instruments.} \label{fig:memstrengthplot}
\end{figure}

All processes have vanishing memory strength for the completely-noisy instrument, which can be implemented by applying random unitaries sampled from a set whose average is the depolarizing channel, providing a convenient way to bound memory~\cite{White2020}. For cases i) and ii), in Fig.~\ref{fig:memstrengthplot} we plot the error in the multi-time expectation value (i.e., l.h.s. of Thm.~\ref{thm:recovery}) and a memory strength proxy based on the relative entropy between the Choi states of the true and recovered processes (see Appendix~\ref{app:proxies}) which upper bounds the r.h.s. of Thm.~\ref{thm:recovery}. The observable $C$ is chosen as an initial preparation, followed by doing nothing for four steps, before a final measurement. Each process displays significant memory strength for the identity instrument, indicating that unperturbed memory does not decay rapidly. In contrast, the effects of interventions are seen for the causal break: here, all memory effects detected result from environmental interactions since the causal break ensures no temporal correlations can be transmitted through the system (cf. the identity instrument). The {CP}-divisible process exhibits negligible memory strength, the intermediate process some, and the strongly non-Markovian one stronger still. We emphasize that the unperturbed evolution is better approximated by the process recovered from the informationally-complete recovery scheme than that from the identity instrument, demonstrating Thm.~\ref{thm:ic}. 


\section*{Discussion}

We have introduced the concept of memory strength for quantum stochastic processes, which is shown to bound process recoverability. Its applicability is exemplified by the case study, where we are able to accurately and efficiently reconstruct dynamics with a memory cutoff, even in a highly non-Markovian regime. We expect these tools to be broadly applicable to modern techniques for efficient simulation, where operationally-motivated memory approximations with quantitative error bounds are desired, such as transfer tensor~\cite{Cerrillo2014, Rosenbach2016, Kananenka2016, Pollock2018T, Jorgensen2019} and machine-learning methods that either attempt to learn non-Markovian features~\cite{Banchi2018, costa2019} or compress the memory to low-dimensional effective environments~\cite{Luchnikov2019, guochu2020}. Our notion of memory strength and the associated concept of recoverability will play an important role in the characterization and mitigation of noise in quantum experiments where multi-time memory effects are present~\cite{Taranto2020Exp, White2020,White2021,White2021Diagnosing}. Of particular relevance in this direction, in Ref.~\cite{White2020} the authors recovered the restricted process tensors on four IBM quantum computers and reported a reconstruction fidelity of order $10^{-3}$. Additionally, in Refs.~\cite{White2021,White2021Diagnosing} the authors directly applied our tools to drastically reduce the number of conditional circuits required to be estimated to characterize a multi-time process on the IBM quantum computer. In particular, they demonstrate that a finite Markov order model of length $\ell = 2, 3$ suffices to prepare future target states on a five step non-Markovian process with approximately $88\%$ and $93\%$ fidelity, respectively, in the presence of correlated noise. This is a huge improvement over previous `gold standard' techniques using gate-set tomography to mitigate state-preparation-and-measurement errors, which gives fidelity values of around $75\%$ in the same circumstances. Our present work lays the conceptual foundations for approximating processes with finite memory with requirements only as large as the complexity of the memory and these examples highlight the efficacy of our framework in realistic settings. Further developments in this direction will bridge the gap between efficient characterization and simulation of quantum processes with memory~\cite{Luchnikov2018, Luchnikov2019L,Taranto2020Exp, White2020,White2021,White2021Diagnosing}. 


\section*{Methods}
\setcounter{subsection}{0}

\subsection{Introduction to Process Tensor}

Here we provide an introduction to the process tensor formalism; for details, see, e.g., Refs.~\cite{Chiribella2009,Pollock2018A,Milz2017,Milz2021}. 

A discrete-time classical stochastic process is characterized by the joint probability distribution $\mathbbm{P}$ over all sequences of events, $\mathbbm{P}( x_n, \hdots, x_1)$, where we drop the explicit time labels with the understanding that $x_j$ represents an event at time $t_j$. In multi-time quantum processes, it is important to not only capture the outcome of a measurement, but also the transformation induced on the state, which together constitute an event. Thus, an interrogation of a quantum stochastic process at time $t_j$ is described by an instrument $\mathcal{J}_j = \{ \mathcal{O}_j^{(x_j)} \}$, which is a collection of completely positive (CP) maps that sum to a completely positive and trace preserving (CPTP) map. Instruments represent general quantum operations, including projective measurements, unitary transformations, and anything in between. Each CP map corresponds to a particular event realized and the fact that the maps sum to a CPTP one encodes the assumption that some event is observed. A discrete-time quantum stochastic process is uniquely described once the probability $\mathbbm{P}(x_n, \hdots, x_1 | \mathcal{J}_n, \hdots, \mathcal{J}_1)$ for all possible events $\{ x_1, \hdots, x_n \}$ for all possible instruments $\{ \mathcal{J}_1, \hdots, \mathcal{J}_n\}$ are known. As a consequence of the linearity of mixing principle~\cite{ShrapnelCosta2017}, there exists a multi-linear functional $\mathcal{T}_{n:1}$ that takes any sequence of CP maps to the correct probability distribution via $\mathbbm{P}({x_n, \hdots, x_1}|\mathcal{J}_n, \hdots, \mathcal{J}_1) = \mathcal{T}_{n:1}[\mathcal{O}_n^{(x_n)}, \hdots, \mathcal{O}_1^{(x_1)}]$, known as the process tensor~\cite{Pollock2018A}. The process tensor generalizes classical stochastic processes~\cite{Milz2017KET} and reproduces classical properties appropriately~\cite{strasberg_classical_2019,Milz2020}. As it encodes all detectable memory effects, it has been used to develop operationally meaningful notions of quantum Markovianity~\cite{Pollock2018L, Pollock2018A} and memory length~\cite{Taranto2019L, Taranto2019A, TarantoThesis}. 

Since all of the CP maps constituting the instruments, as well as the process tensor itself, are linear maps, they can be represented as matrices through the Choi-Jamio{\l}kowski isomorphism (CJI)~\cite{Milz2017}. Any map $\mathcal{O}: \mathcal{B}(\mathcal{H}_\inp) \to \mathcal{B}(\mathcal{H}_\out)$, where $\mathcal{B}(\mathcal{X})$ denotes the space of bounded linear operators on $\mathcal{X}$, can be mapped isomorphically to a matrix $\mathsf{O} \in \mathcal{B}(\mathcal{H}_\out \otimes \mathcal{H}_\inp)$ through its action on half of an (unnormalized) maximally entangled state $\Psi^+ := \sum_{i,j} \ket{ii}\bra{jj} \in \mathcal{B}(\mathcal{H}_\inp \otimes \mathcal{H}_\inp)$, i.e., $\mathsf{O} := (\mathcal{O} \otimes \mathcal{I})[\Psi^+]$. Note that the time of the event is associated to both an input and output Hilbert space, and the Choi matrix is a supernormalized bipartite state. In this representation, the properties of CP and TP for the maps respectively translate to $\mathsf{O} \geq 0$ and $\ptr{\out}{\mathsf{O}} = \mathbbm{1}_\inp$. To aid intuition, note that the output state $\rho^\prime := \mathcal{O}[\rho]$ of the map $\mathcal{O}$ acting on an arbitrary input state $\rho$ is computed in the Choi picture via $\rho^\prime = \ptr{\inp}{(\mathbbm{1}_\out \otimes \rho_\inp^{\textup{T}}) \mathsf{O}}$. If the initial state is not subject to a (CPTP) quantum channel but instead a particular event is observed, associated to a CP map of an instrument $\mathcal{J} = \{ \mathsf{O}^{(x)} \}$, then the corresponding probability is computed via
\begin{align}\label{eq:choistatemapmeasure}
    \mathbbm{P}(x|\mathcal{J}) = \tr{(\mathbbm{1}_\out \otimes \rho^{\textup{T}}_{\inp}) \mathsf{O}^{(x)}_{\out \inp}}.
\end{align}

Similarly, the action of a process tensor map $\mathcal{T}_{n:1}$ on a sequence of instrument elements $\{ \mathcal{O}_1^{(x_1)}, \hdots, \mathcal{O}_{n}^{(x_n)}\}$ can be expressed in terms of a multiplication of their Choi matrices and a trace as follows~\cite{ShrapnelCosta2017}
\begin{gather}\label{eq:appgbornrule}
\begin{split}
    &\mathbbm{P}(x_n, \hdots, x_1 | \mathcal{J}_n, \hdots, \mathcal{J}_1) \\
    &= \tr{(\mathsf{O}_n^{(x_n)} \otimes \hdots \otimes \mathsf{O}_1^{(x_1)} )^\textup{T} \Upsilon_{n:1} },
\end{split}    
\end{gather}
where $\Upsilon_{n:1} \in \mathcal{B}(\mathcal{H}_{n^\inp} \otimes \mathcal{H}_{n-1^\out} \otimes \hdots \otimes \mathcal{H}_{1^\inp})$ is the $(2n-1)$-partite Choi matrix of the process tensor map $\mathcal{T}_{n:1}$ and each $\mathsf{O}_j^{(x_j)}$ is the Choi matrix of $\mathcal{O}_j^{(x_j)}$. The Choi state of the process $\Upsilon_{n:1}$ essentially plays the role of a quantum state over time, insasmuch as it encodes all observable probability distributions for all possible instrument sequences (just as a quantum state encodes all observable probability distributions for any choice of POVM). In the (one-time) spatial setting, $\Upsilon_1 = \mathbbm{1}_\out \otimes \rho^{\textup{T}}_{\inp}$ and Eq.~\eqref{eq:appgbornrule} reduces to Eq.~\eqref{eq:choistatemapmeasure}. Note that we label the Hilbert spaces logically from the perspective of the experimenter (i.e., the experimenter receives a state from the process that is ``input'' into their instrument of choice, transforming it into an ``output'' state that is fed back into the process); hence, $\inp$ denotes outputs of the process and $\out$ denotes inputs to the process. Thus, whenever a process tensor acts on an instrument sequence, the degrees of freedom with the same labels (timestep and input/output) are contracted over. 

The natural generalisation $\Upsilon_{n:1}$ of the CJI applied to the multilinear map $\mathcal{T}_{n:1}$ is constructed by feeding one half of an (unnormalized) maximally entangled state into the dynamics at each time~\cite{Milz2017}. More precisely, begin with the system-environment dilated dynamics shown in Fig.~\ref{fig:processtensor} (green), and denote the initial system-environment state by $\rho$ and the unitary maps describing the joint evolution between times $t_{j-1}$ and $t_j$ by $\mathcal{U}_{j:j-1}$. Now consider $n-1$ additional maximally entangled pairs, $\Psi^+_{j^\out j^\out}$ associated to auxilliary systems $A_{j^\out}\simeq S$, collectively described as $\Psi^{+}_{n-1} := \bigotimes_{j=1}^{n-1}  \Psi^+_{j^\out j^\out}$. Letting the unitary maps between each timestep act on the environment and one half of the appropriate auxilliary systems, i.e., $\mathcal{U}_{j:j-1}: \mathcal{B}(\mathcal{H}_{j-1^\out} \otimes \mathcal{H}_E) \to \mathcal{B}(\mathcal{H}_{j^\inp} \otimes \mathcal{H}_E)$ yields the Choi state of the process tensor:
\begin{align}\label{eq:cjiprocesstensor}
    \Upsilon_{n:1} =&\, \mathrm{tr}_E[\mathcal{U}_{n:n-1}  \hdots \mathcal{U}_{2:1}(\rho \otimes \Psi^{+}_{n-1})].
\end{align}
It is straightforward (albeit arduous) to verify the correctness of Eq.~\eqref{eq:appgbornrule} via direct insertion of Eq.~\eqref{eq:cjiprocesstensor}. Natural generalizations of complete positivity and trace preservation to multi-time processes translate respectively to $\Upsilon_{n:1} \geq 0$ and the following hierarchy of trace conditions:
\begin{align}\label{eq:causalityhierarchy}
    \ptr{j^\inp}{\Upsilon_{j:1}} = \mathbbm{1}_{j-1^\out} \otimes \Upsilon_{j-1:1}, \quad \forall j.
\end{align}
Conversely, any operator satisfying the above represents some (causally-ordered) quantum dynamics inasmuch as it corresponds to an underlying system-environment circuit~\cite{Chiribella2009,Pollock2018A}.

Eq.~\eqref{eq:appgbornrule} constitutes a special case of how (higher-order) quantum maps act on each other: Here, we are contracting all open slots of the process tensor with an operation associated to each time in order to yield a probability distribution. More generally, it is possible to consider applying instruments to only a subset of times, yielding a ``conditional'' process defined upon the remaining times, which describes the correct behaviour of the concatenated dynamics. In other words, it contains all of the information required to compute the correct probability distribution for any instruments subsequently applied to the remaining times. To compute such an object in the Choi representation, one uses the \emph{link product} defined in Ref.~\cite{Chiribella2008-2}. Essentially, this amounts to restricting both the trace and the transposition in Eq.~\eqref{eq:appgbornrule} to only the common Hilbert spaces associated to the relevant subset of times where the instrument is being applied. 

For instance, grouping the times into history $\{ t_1, \hdots, t_{k}\}$, memory $\{ t_{k+1}, \hdots , t_{k+\ell}\}$ and future $\{ t_{k+\ell+1}, \hdots, t_n\}$ and choosing an instrument $\mathcal{J}_M = \{ \mathsf{O}_M^{(x_M)} \}$ on the memory block, the conditional future-history process that occurs given any particular event sequence $\mathsf{O}_M^{(x_M)}$ on $M$ alone is
\begin{align}
    \widetilde{\Upsilon}_{FH}^{(x_M)} = \ptr{M}{\mathsf{O}_M^{(x_M) \textup{T}} \Upsilon_{FMH}}.
\end{align}
Such a conditional process is generically correlated across $F$ and $H$; however, if it is of tensor product form $\widetilde{\Upsilon}_{FH}^{(x_M)} = \Upsilon_F^{(x_M)} \otimes \widetilde{\Upsilon}_H^{(x_M)}$ for each event $x_M$ of the instrument $\mathcal{J}_M$, the process has Markov order $\ell := |M|$ with respect to said instrument~\cite{Taranto2019L}. (In general, each $\mathsf{O}^{(x_M)}_M$ may act on only a subspace of $M$, with the history and future retaining the rest, to yield {$\widetilde{\Upsilon}_{FM_F M_H H}^{(x_M)}$}, where {$M_F$} and {$M_H$} can depend on {$x_M$} [see Appendix~\ref{app:qmoexample}]; for brevity, we absorb these into {$F$} and {$H$}.)

Lastly, a Markovian process corresponds to one for which the process tensor has the specific tensor product structure of an uncorrelated sequence of CPTP maps $\{ \Lambda_{j^\inp:j-1^\out}\}$ connecting adjacent timesteps, and an initial quantum state $\rho_{1^\inp}$~\cite{Pollock2018L}:
\begin{align}
    \Upsilon_{n:1}^{\textup{Markov}} = \Lambda_{n^\inp:n-1^\out} \otimes \Lambda_{n-1^\inp:n-2^\out} \otimes \Lambda_{2^\inp:1^\out} \otimes \rho_{1^\inp}.
\end{align}

\subsection{Preliminaries}

To begin with, we introduce the following definition:
\begin{defn}[Instrument relative entropy]\label{def:instrumentrelativeentropy}
For any family of instruments $\mathbb{J}$ and process tensors $\Upsilon$, $\Gamma$,
\begin{gather}
    S_{\mathbb{J}}(\Upsilon\|\Gamma):= \sup_{\mathcal{J}\in \mathbb{J}} S\left(\mathcal{P}_\mathcal{J}[\Upsilon]\middle\|\mathcal{P}_\mathcal{J}[\Gamma]\right),
\end{gather}
where $S(A\|B):= \tr{A(\log A - \log B)}$ is the usual quantum relative entropy and $\mathcal{P}_\mathcal{J}$ is a CP map from process tensors to classical pointer states, whose elements form a probability distribution over outcomes of the instrument $\mathcal{J} = \{\mathsf{O}^{(x)}\}$:
\begin{gather}
    \mathcal{P}_\mathcal{J}[\Upsilon]:= \sum_{x} \tr{\mathsf{O}^{(x) \textup{T}} \Upsilon} \ketbra{x}{x}.
\end{gather}
\end{defn}

\begin{prop} \label{prop:loosethetabound}
For any $\Upsilon_{FMH}$, $\mathcal{J}_M$ and $\uln{\Lambda}^{\mathcal{J}_M}_{FMH}$ as defined in Eq.~\eqref{eq:restricted},
\begin{gather} \label{eq:loosethetabound}
    \Theta(\mathcal{J}_M) = S_{\mathbb{J}\cap {\rm span}(\mathcal{J}_M)}\left(\Upsilon_{FMH}\middle\|\uln{\Lambda}^{\mathcal{J}_M}_{FMH}\right),
\end{gather}
with $\Theta(\mathcal{J}_M)$ taken to be the measured CMI, $\mathbb{J}\cap {\rm span}(\mathcal{J}_M)$ a family of instruments whose elements have support on $M$ only in the linear span of the elements of $\mathcal{J}_M$.
\end{prop}
\noindent\textit{Proof.} Consider first the measured conditional probability distributions for a fixed memory instrument $\mathcal{J}_M$ and arbitrary (independent) $\mathcal{J}_F,\mathcal{J}_H$ arising from the process tensor $\Upsilon_{FMH}$ and the recovered process $\uln{\Lambda}_{FMH}^{\mathcal{J}_M}$, which are respectively given by:
\begin{align}
    \mathbbm{P}_{\Upsilon_{FMH}}(x_F,x_H|x_M) \!=\!\tr{(\mathsf{O}_F^{(x_F)} \otimes \mathsf{O}_H^{(x_H)})^\textup{T} \widetilde{\Upsilon}_{FH}^{(x_M)}}\!
\end{align}
and 
\begin{align}
    \mathbbm{P}_{\uln{\Lambda}_{FMH}^{\mathcal{J}_M}}(x_F,x_H|x_M)\!=\!\tr{(\mathsf{O}_F^{(x_F)} \otimes \mathsf{O}_H^{(x_H)})^\textup{T} \widetilde{\uln{\Lambda}}_{FH}^{(x_M)}}\!.
\end{align}
Here, we have defined $\widetilde{\uln{\Lambda}}_{FH}^{(x_M)} := \ptr{M}{\mathsf{O}_M^{(x_M) \textup{T}} \uln{\Lambda}_{FMH}^{\mathcal{J}_M}}$, which, by construction, factorizes as
\begin{align}
    \mathbbm{P}_{\uln{\Lambda}_{FMH}^{\mathcal{J}_M}}&(x_H,x_H|x_M) \notag \\ 
    &=\tr{(\mathsf{O}_F^{(x_F)} \otimes \mathsf{O}_H^{(x_H)})^\textup{T} (\Upsilon_F^{(x_F)} \otimes \widetilde{\Upsilon}_{H}^{(x_H)}) } \notag \\
    &= \mathbbm{P}_{\Upsilon_{FMH}}(x_F|x_M) \mathbbm{P}_{\Upsilon_{FMH}}(x_H|x_M), 
\end{align}
since $\widetilde{\uln{\Lambda}}_{FH}^{(x_M)} = \Upsilon_F^{(x_F)} \otimes \widetilde{\Upsilon}_{H}^{(x_H)}$.

With this, we can express the mutual information in the correlated distribution $I(F:H)_{\mathbbm{P}_{\Upsilon}(x_F,x_H|x_M)}$ in terms of the relative entropy between said distribution and the uncorrelated one arising from measurements on the recovered process, i.e.,
\begin{align}
    I_{\mathbbm{J}}&(F:H)_{\Upsilon} = \notag \\ &S[\mathbbm{P}_{\Upsilon_{FMH}}(x_F,x_H|x_M) \| \mathbbm{P}_{\uln{\Lambda}_{FMH}^{\mathcal{J}_M}}(x_F,x_H|x_M)].
\end{align}
Thus, beginning with the definition of Eq.~\eqref{eq:instrumentstrength}, we have:
\begin{align}
    \Theta(\mathcal{J}_M) &= \sup_{\mathcal{J}_F,\mathcal{J}_H \in \mathbbm{J}} \left( \sum_{x_M} p(x_M|\mathcal{J}_M)I_{\mathbbm{J}}(F:H|x_M)_\Upsilon \right) \notag \\
    &= S_{\mathbbm{J}_{FmH}}(\Upsilon_{FmH}^{\mathcal{J}_M} \| \Lambda_{FmH}^{\mathcal{J}_M}).
\end{align}
Here $\mathbbm{J}_{FmH}$ is the original set of uncorrelated instruments $\mathbbm{J}$ on $FH$, combined with a POVM on the pointer space $m$, i.e., the supremum in the measured relative entropy is taken over $\mathcal{J} = \{\mathsf{O}^{(x)}_{FmH}\} \in \mathbbm{J}_{FmH}$,
\begin{gather}
\mathsf{O}^{(x)}_{FmH} = \sum_{x_M} \mathsf{E}^{(x, x_M)}_{FH}\otimes \ketbra{x_M}{x_M}_m,    
\end{gather}
where $\mathsf{E}^{(x, x_M)}_{FH}$ can be any operator, $\mathsf{O}^\mathcal{J}_{FmH} =\sum_{x} \mathsf{O}^{(x)}_{FmH}$ satisfies the relevant trace conditions on the $FH$ part and $\ptr{FH}{\mathsf{O}^\mathcal{J}_{FmH}} = D^\mathtt{i}_{FH}\mathbbm{1}_m$. Since, for $\mathcal{J}_M = \{\mathsf{O}^{(x_M)}_M\}$,
\begin{gather}
    \ptr{m}{\ketbra{x_M}{x_M}_m {\Gamma}^{\mathcal{J}_M}_{FmH}} = \ptr{M}{\mathsf{O}_M^{(x_M)} \sum_{y_M} \Gamma^{(y_M)}_{FH} \otimes \mathsf{D}^{(y_M)}_M},\nonumber
\end{gather}
with $\Gamma \in \{\Upsilon, \Lambda\}$ and $\{\mathsf{D}^{(x_M)}_M\}$ the dual set to $\mathcal{J}_M$, we have
\begin{gather}
S_{\mathbbm{J}_{FmH}}({\Upsilon}^{\mathcal{J}_M}_{FmH}\|{\Lambda}_{FmH}^{\mathcal{J}_M}) =  S_{\mathbb{J}\cap {\rm span}(\mathcal{J}_M)}(\Upsilon_{FMH}\|\uln{\Lambda}^{\mathcal{J}_M}_{FMH}),   \notag 
\end{gather}
with $\uln{\Lambda}^{\mathcal{J}_M}_{FMH} = \sum_{x_M} \Upsilon^{(x_M)}_{F} \otimes \mathsf{D}^{(x_M)}_M \otimes \widetilde{\Upsilon}^{(x_M)}_{H}$. Hence, the claim is asserted. \hfill $\square$

We are now in a position to prove our main results. 

\subsection{Proof of Thm.~\ref{thm:recovery}}\label{app:thmrecovery}

First, we note that, for any set of instruments $\mathbb{J}$ and process tensors $\Upsilon$ and $\Gamma$,
\begin{gather}
    S_{\mathbb{J}}(\Upsilon\|\Gamma) = \sup_{\mathcal{J} \in \mathbb{J}}\sum_x p^\mathcal{J}_x(\log p^\mathcal{J}_x - \log q^\mathcal{J}_x),
\end{gather}
with $p^\mathcal{J}_x := \tr{\mathsf{O}^{(x) \textup{T}}\Upsilon_{FMH}}$ and $q^\mathcal{J}_x := \tr{\mathsf{O}^{(x)\textup{T}}\uln{\Lambda}^{\mathcal{J}_M}_{FMH}}$ the probabilities associated with the instrument $\mathcal{J} = \{\mathsf{O}^{(x)}\}$. We can then use Pinsker's inequality to write 
\begin{align}\label{eq:instrumententropybound}
    S_{\mathbb{J}}(\Upsilon\|\Gamma) \geq& \frac{1}{2}\sup_{\mathcal{J} \in \mathbb{J}}\left(\textstyle{\sum_x} |p^\mathcal{J}_x-q^\mathcal{J}_x|\right)^2  \\
    =& \frac{1}{2}\left(\sup_{\mathcal{J} \in \mathbb{J}}\textstyle{\sum_x} \left|\tr{\mathsf{O}^{(x)\textup{T}}(\Upsilon - \Gamma)}\right|\right)^2. \nonumber
\end{align}

Any multi-time operator $C$ can be decomposed in terms of the elements of a single instrument $\mathcal{J} = \{{\mathsf{O}}^{(x)}\}\in\mathbb{J}$, as long as those elements span a sufficiently large space. That is, $C = \sum_{x} c^\mathcal{J}_x {\mathsf{O}}^{(x)}$ with $c^\mathcal{J}_x\in\mathbb{C}$; in general, the norm $|C|_\mathcal{J}:=\sqrt{\sum_x |c^\mathcal{J}_x|^2}$ will vary with the instrument involved in the decomposition. We therefore have, using the Cauchy-Schwarz inequality,
\begin{align} \label{eq:cgammacauchyschwarz}
    \left|\tr{C\Xi}\right| 
    \leq& \inf_{\mathcal{J}\in \mathbb{J}}\left(\sqrt{\textstyle{\sum_x} |c_x^\mathcal{J}|^2} \sqrt{\textstyle{\sum_x} \left|\tr{{\mathsf{O}}^{(x)\textup{T}} \Xi}\right|^2}\right) \nonumber \\
    \leq& \inf_{\mathcal{J}\in \mathbb{J}}\left(|C|_\mathcal{J}\textstyle{\sum_x} \left|\tr{{\mathsf{O}}^{(x)\textup{T}} \Xi}\right|\right) \nonumber\\
    \leq&  \inf_{\mathcal{J} \in \mathbb{J}}\left(|C|_\mathcal{J}\right)\sup_{\mathcal{J} \in \mathbb{J}}\left(\textstyle{\sum_x} \left|\tr{\mathsf{O}^{(x)\textup{T}}\Xi}\right|\right), 
\end{align}
for any operator $\Xi$. Choosing $\Xi = \Upsilon_{FMH} - \uln{\Lambda}^{\mathcal{J}_M}_{FMH}$ and restricting the set $\mathbb{J}$ to $\mathbb{J}\cap {\rm span}(\mathcal{J}_M)$, such that $\tr{C\Xi}=\langle C \rangle_{\Upsilon_{FMH}} - \langle C \rangle_{\uln{\Lambda}^{\mathcal{J}_M}_{FMH}}$, and combining Eqs.~\eqref{eq:instrumententropybound}~and~\eqref{eq:cgammacauchyschwarz} leads to the bound
\begin{align} \label{eq:Cbound}
    & \left|\langle C \rangle_{\Upsilon_{FMH}} - \langle C \rangle_{\uln{\Lambda}^{\mathcal{J}_M}_{FMH}}\right| \nonumber \\
    \leq& |\mathbf{C}| \sqrt{2S_{\mathbb{J}\cap {\rm span}(\mathcal{J}_M)}(\Upsilon_{FMH}\|\uln{\Lambda}^{\mathcal{J}_M}_{FMH})},
\end{align}
where $|\mathbf{C}|=\min_{\mathcal{J}\in\mathbb{J}\cap {\rm span}(\mathcal{J}_M)} |C|_\mathcal{J}$ (equivalent to the definition given in Thm.~\ref{thm:recovery}). Invoking Prop.~\ref{prop:loosethetabound} proves Thm.~\ref{thm:recovery}. \hfill $\square$

Cor.~\ref{cor:densityop} follows directly as shown below.

\subsection{Proof of Cor.~\ref{cor:densityop}}\label{app:cor2}

Choose $C = \mathsf{P}_j \otimes {\Psi^+}^{\otimes j-k-2} \otimes \mathsf{O}^{(x_M)}_M \otimes {\Psi^+}^{\otimes k-\ell-2}$, with $\Psi^+ := \sum_{\alpha \beta} \ket{\alpha \alpha}\bra{\beta \beta}$ the Choi state of the identity map and $\mathsf{P}_j$ a projector. Then $|\mathbf{C}| = 1$, since $C$ is an element of the instrument where the system is left to freely evolve on $H$, $\mathcal{J}_M$ is applied, it again freely evolves to time $t_j$ and then the POVM $\{\mathsf{P}_j, \mathbbm{1}-\mathsf{P}_j\}$ is applied. It follows that $\langle C \rangle_\Upsilon = \tr{\mathsf{P}_j \rho_j^{(x_M)}}$, with $\rho_j^{(x_M)} := \ptr{FMH\backslash j}{\left(\Psi^+_H \otimes \mathsf{O}_M^{(x_M)} \otimes \Psi^+_{F\backslash j}\right) \Upsilon_{FMH}}$ the state, at time $t_j$, of the system undergoing the process specified by $\Upsilon_{FMH}$, acted on by $\mathcal{J}_M$ with outcome $x_M$ occurring, and with no other active interventions. Similarly, the predicted state is $\rho_j^{\prime (x_M)} := \ptr{FMH\backslash j}{\left(\Psi^+_H \otimes \mathsf{O}_M^{(x_M)} \otimes \Psi^+_{F\backslash j}\right) \uln{\Lambda}^{\mathcal{J}_M}_{FMH}}$. Here, $\Psi^+$ denotes the Choi state of the identity map and $\Psi^+_X$ is shorthand for a sequence of identity maps applied at all times in the block $X$, and $j$ corresponds to a subspace of the future Hilbert space associated to time $t_j$. The l.h.s. of Eq.~\eqref{eq:theorem} of the main text then reduces to $|\mathrm{tr}[\mathsf{P}_j(\rho_j^{(x_M)}-{\rho'}_j^{(x_M)})]|$; since the bound must be true for any $\mathsf{P}_j$, it must be true for the one for which the l.h.s. is largest; i.e., $\sup_{\mathsf{P}_j}|\mathrm{tr}[\mathsf{P}_j(\rho_j^{(x_M)}-{\rho'}_j^{(x_M)})]| = \|\rho_j^{(x_M)}-{\rho'}_j^{(x_M)}\|_1$ is bounded. \hfill $\square$

For informationally complete instruments, a combination of the results derived above leads to Thm.~\ref{thm:ic}.

\subsection{Proof of Thm.~\ref{thm:ic}}\label{app:thmic}

When $\mathcal{J}_M$ is informationally complete, $\uln{\Lambda}^{\mathcal{J}_M}_{FMH}$ is a full process tensor and any instrument can be applied to it, since
$\mathbb{J}\cap {\rm span}(\mathcal{J}_M) = \mathbb{J}$ by definition. Therefore, we can use Eq.~\eqref{eq:instrumententropybound}, along with Prop.~\ref{prop:loosethetabound}, to write:
\begin{gather}
    \left(\sup_{\mathcal{J} \in \mathbb{J}}\textstyle{\sum_x} \left|\tr{\mathsf{O}^{(x)}(\Upsilon_{FMH} - \uln{\Lambda}^{\mathcal{J}_M}_{FMH})}\right|\right)^2 \leq 2  \Theta(\mathcal{J}_M)
\end{gather}
The square root of the l.h.s. of this equation is the generalized diamond distance $\left\|\Upsilon_{FMH}-\uln{\Lambda}^{\mathcal{J}_M}_{FMH}\right\|_\diamond$ with $\|X\|_\diamond$ defined in Thm.~\ref{thm:ic}. Equation~\eqref{eq:diamonddistance} and hence Thm.~\ref{thm:ic} follows. \hfill $\square$


\begin{acknowledgments}
We thank Top Notoh, Joshua Morris and Simon Milz for discussions. P. T. was supported by the Australian Government Research Training Program Scholarship, the J. L. William Scholarship, and the Austrian Science Fund (FWF): Y879-N27 (START project). K. M. is supported through the Australian Research Council Future Fellowship FT160100073. This research was supported in part by the National Science Foundation under Grant No. NSF PHY-1748958.
\end{acknowledgments}



\section*{Author Contributions}

All authors contributed to conception of the project, proved and interpreted the results, and discussed and edited the manuscript. P. T. wrote the text and performed all numerical simulations for the case study.



%

\appendix 
\section*{APPENDICES}

\section{Memory Effects in Shallow Pocket Model}\label{app:shallowpocket}

Here we detail the shallow pocket model considered as a motivating example~\cite{Lindblad1980, Accardi1982,Arenz2015,Arenz2017}. It describes a qubit system coupled to a linear degree of freedom that acts as its environment. The dynamics is generated by the interaction Hamiltonian
\begin{align}
H = \frac{g}{2} \sigma_z \otimes \hat{x} = \frac{g}{2} \left[ \begin{array}{cc} x & 0 \\
0 & -x \end{array}\right],
\end{align}
where $\hat{x}$ is the position operator and $g$ the coupling strength. The joint state of the system-environment evolves as
\begin{align}
\rho^{SE}_t (x) &= \left[ \begin{array}{cc} \rho^{11}_0(x) & \rho^{10}_0(x)\textup{e}^{igxt} \\
\rho^{01}_0(x)\textup{e}^{-igxt} & \rho^{00}_0(x) \end{array}\right],
\end{align}
where $\rho^{ij}_0$ corresponds to the matrix element $(i,j)$ of the system density operator at time $t=0$.

In order to examine memory effects in this process, we consider the initial state of the system to be maximally entangled with an additional ancilla $A$. The environment begins uncorrelated with $SA$ in the state $\ket{\psi}\bra{\psi}^E$, which is such that
\begin{align}
	\braket{x|\psi} = \sqrt{\frac{\gamma}{\pi}} \frac{1}{x+i \gamma}. 
\end{align}
We track the mutual information $I(S:A) := S(\rho^S)+S(\rho^A) - S(\rho^{SA})$ as the system is subject to this dynamics. 

Firstly, suppose that no interventions are made to the system as it evolves. Throughout the process, the system builds up correlations with the environment at the expense of those shared with the ancilla. This can be seen by tracing out the environmental degrees of freedom: in doing so, one obtains---through the inverse Fourier transform of a Lorentzian distribution---the following system-ancilla time-evolved state
\begin{align}
	\rho^{SA}_t = \frac{1}{2} \left[ \begin{array}{cccc}
	1 & 0 & 0 & \textup{e}^{- g \gamma t} \\
	0 & 0 & 0 & 0 \\
	0 & 0 & 0 & 0 \\
	\textup{e}^{- g \gamma t} & 0 & 0 & 1 \end{array}\right].
\end{align}
The mutual information of this state decays exponentially in time, as depicted by the black, solid curve in Fig.~\ref{fig:shallowpocket} of the main text (note that we choose $g=0.8, \gamma=0.3$ for all curves in this figure).

We now examine the effect of implementing an operation on the system at some intermediary time. We consider the natural evolution above to occur up until some fixed time $t_1$ ($= 5$ for illustration), at which point an arbitrary quantum operation can be applied to the system; the system subsequently evolves according to the shallow pocket model up to some later time $t_1 + \tau$. It is crucial to track the entire system-ancilla-environment state throughout the process to understand the evolution of the correlations between the parties---as this is how memory effects are made manifest---with the environment only being discarded at the conclusion. 

Consider first applying the Pauli rotation $\sigma_x$ to the system at $t_1$. The subsequent joint system-ancilla state at $t_1 + \tau$ is
\begin{align}
	\rho_{t_1+\tau}^{SA}(\sigma_x) = \frac{1}{2} \left[ \begin{array}{cccc}
	0 & 0 & 0 & 0 \\
	0 & 1 & \textup{e}^{- g \gamma |t_1 - \tau|} & 0 \\
	0 & \textup{e}^{- g \gamma |t_1 - \tau|} & 1 & 0 \\
	0 & 0 & 0 & 0 \end{array}\right].
\end{align}
where the notation means that $\sigma_x$ was applied at fixed time $t_1$, followed by shallow pocket evolution for variable time $\tau$. It is clear that by time $\tau = t_1$, the system-ancilla has returned to a maximally correlated state. The mutual information as this state evolves is depicted by the blue, long-dashed line in Fig.~\ref{fig:shallowpocket} of the main text. Indeed, this analysis recovers the well-known result that application of $\sigma_x$ at time $t_1$ reverses the dynamics and leads to the system returning to its initial state (of maximal correlation with the ancilla, in our case) by time $2 t_1$~\cite{Arenz2015,Arenz2017}. This follows directly from the identity $\sigma_x H \sigma_x^\dagger = -H$, which leads to $\textup{e}^{i t_1 H} \sigma_x \textup{e}^{i t_1 H} \sigma_x^\dagger = \mathbbm{1}$.

One can also consider the experimenter applying other operations. For instance, perhaps the operation implemented at $t_1$ is some offset Pauli rotation $\sigma_{\textup{offset}} := \sqrt{p} \sigma_x + \sqrt{1-p}\sigma_z$. In this case, the subsequent system-ancilla state is 
\begin{widetext}
\begin{align}
	\rho_{t_1+\tau}^{SA}(\sigma_{\textup{offset}} ) = \frac{1}{2} \left[ \begin{array}{cccc}
	1-p & \sqrt{p (1-p)} \textup{e}^{-g \gamma t_1} & \sqrt{p (1-p)}\textup{e}^{-g \gamma \tau} & (1-p) \textup{e}^{- g \gamma (t_1+\tau)} \\
	\sqrt{p (1-p)} \textup{e}^{-g \gamma t_1} & p & p\textup{e}^{- g \gamma |t_1 - \tau|} & -\sqrt{p (1-p)} \textup{e}^{-g \gamma \tau} \\
	\sqrt{p (1-p)} \textup{e}^{-g \gamma \tau} & p\textup{e}^{- g \gamma |t_1 - \tau|} & p & -\sqrt{p (1-p)} \textup{e}^{-g \gamma t_1} \\
	 (1-p) \textup{e}^{- g \gamma (t_1+\tau)} & -\sqrt{p (1-p)} \textup{e}^{-g \gamma \tau} & -\sqrt{p (1-p)} \textup{e}^{-g \gamma t_1} & 1-p \end{array}\right].
\end{align}
\end{widetext}
The mutual information of this state is depicted by the green dotted line in Fig.~\ref{fig:shallowpocket} of the main text. Interestingly, this first induces a decrease in the mutual information between the system and ancilla that is steeper than the exponential decay that occurs when no operation is implemented, before correlations build back up as the system evolves with the environment, which retains memory of the system's past.

Similarly, a measurement of the system state could be made. In Fig.~\ref{fig:shallowpocket} of the main text, the red, dot-dashed curve depicts this for a measurement in the $x$-basis spanned by $\{ \ket{\pm} := \ket{0}\pm\ket{1}\}$; here we show the mutual information for a measurement yielding the outcome $+$. The post-measurement system-ancilla state is
\begin{widetext}
\begin{align}
	\rho_{t_1+\tau}^{SA}(+) = \frac{1}{4} 
	\left[ \begin{array}{cccc}
	1 & \textup{e}^{- g \gamma t_1} & \textup{e}^{- g \gamma \tau}  & \textup{e}^{- g \gamma (t_1+\tau)} \\
	\textup{e}^{- g \gamma t_1} & 1 & \textup{e}^{- g \gamma |t_1 - \tau|} &  \textup{e}^{- g \gamma \tau} \\
	 \textup{e}^{- g \gamma \tau} & \textup{e}^{- g \gamma |t_1 - \tau|} & 1 &  \textup{e}^{- g \gamma t_1} \\
	 \textup{e}^{- g \gamma (t_1+\tau)} &  \textup{e}^{- g \gamma \tau} &  \textup{e}^{- g \gamma t_1} & 1 \end{array}\right]. 
\end{align}
\end{widetext}
Directly after the result is observed, i.e., at $\tau = 0$, the mutual information drops to 0 as the system and ancilla are uncorrelated; however, again, correlations build back up as the system evolves with its environment due to memory. 

Lastly, the experimenter could attempt to completely erase any historic information by discarding the system (i.e., measuring without recording the outcomes) and prepare a fixed, known state to feed into the dynamics. The purple, short-dashed curve in Fig.~\ref{fig:shallowpocket} depicts this scenario. For the shallow pocket model considered, it does not matter which state is prepared by the experimenter: the discarding of measurement outcomes destroys any memory of the history and no correlations between the system and ancilla can ever be built up again. Crucially, this is not the case in general and only occurs because the dynamics is CP-divisible. 

All of the information above is encapsulated in the process tensor, as described in the main text. For the shallow pocket model evolving any input state of the system until time $t_1$, at which point any CP operation can be implemented, followed by an additional time $\tau$ of shallow pocket dynamics, the process tensor is as follows
\begin{widetext}
\begin{align}
\Upsilon_{2:0} =
\left[
\begin{array}{cccccccccccccccc}
1 & 0 & 0 & 0 & 0 & \textup{e}^{-g \gamma t_1} & 0 & 0 & 0 & 0 & \textup{e}^{-g \gamma \tau} & 0 & 0 & 0 & 0 & \textup{e}^{-g \gamma (t_1+\tau)} \\
0 & 0 & 0 & 0 & 0 & 0 & 0 & 0 & 0 & 0 & 0 & 0 & 0 & 0 & 0 & 0 \\
0 & 0 & 0 & 0 & 0 & 0 & 0 & 0 & 0 & 0 & 0 & 0 & 0 & 0 & 0 & 0 \\
0 & 0 & 0 & 0 & 0 & 0 & 0 & 0 & 0 & 0 & 0 & 0 & 0 & 0 & 0 & 0 \\
0 & 0 & 0 & 0 & 0 & 0 & 0 & 0 & 0 & 0 & 0 & 0 & 0 & 0 & 0 & 0 \\
\textup{e}^{-g \gamma t_1}  & 0 & 0 & 0 & 0 & 1 & 0 & 0 & 0 & 0 & \textup{e}^{-g \gamma |t_1-\tau|} & 0 & 0 & 0 & 0 & \textup{e}^{-g \gamma \tau} \\
0 & 0 & 0 & 0 & 0 & 0 & 0 & 0 & 0 & 0 & 0 & 0 & 0 & 0 & 0 & 0 \\
0 & 0 & 0 & 0 & 0 & 0 & 0 & 0 & 0 & 0 & 0 & 0 & 0 & 0 & 0 & 0 \\
0 & 0 & 0 & 0 & 0 & 0 & 0 & 0 & 0 & 0 & 0 & 0 & 0 & 0 & 0 & 0 \\
0 & 0 & 0 & 0 & 0 & 0 & 0 & 0 & 0 & 0 & 0 & 0 & 0 & 0 & 0 & 0 \\
\textup{e}^{-g \gamma \tau}  & 0 & 0 & 0 & 0 & \textup{e}^{-g \gamma |t_1-\tau|}  & 0 & 0 & 0 & 0 & 1 & 0 & 0 & 0 & 0 & \textup{e}^{-g \gamma t_1} \\
0 & 0 & 0 & 0 & 0 & 0 & 0 & 0 & 0 & 0 & 0 & 0 & 0 & 0 & 0 & 0 \\
0 & 0 & 0 & 0 & 0 & 0 & 0 & 0 & 0 & 0 & 0 & 0 & 0 & 0 & 0 & 0 \\
0 & 0 & 0 & 0 & 0 & 0 & 0 & 0 & 0 & 0 & 0 & 0 & 0 & 0 & 0 & 0 \\
0 & 0 & 0 & 0 & 0 & 0 & 0 & 0 & 0 & 0 & 0 & 0 & 0 & 0 & 0 & 0 \\
\textup{e}^{-g \gamma (t_1+\tau)} & 0 & 0 & 0 & 0 & \textup{e}^{-g \gamma \tau}  & 0 & 0 & 0 & 0 & \textup{e}^{-g \gamma t_1} & 0 & 0 & 0 & 0 & 1 
\end{array} \right].
\end{align}
\end{widetext}
The representation of the process tensor used above is a supernormalized four-partite quantum state encapsulating all of the information necessary to predict the output quantum system at time $t_1 + \tau =: t_2$ for any initial system state and any operation applied at $t_1$. Thus it includes all possible memory effects in the process. The instrument-specific memory strength introduced in the main text captures all correlations between the input and output state for any operation applied in between; indeed, the above example considering the correlations between the system and an ancilla with which it is initially maximally entangled provides an alternate interpretation (for two-time memory considerations; for more complex memory structures between multiple timesteps, one would need to consider the behavior of correlations between multiple pairs of initially entangled system-ancilla states, with a copy of the system being fed into the process at each timestep).

\section{Process with Finite Quantum Markov Order with Parts of $M$ kept by $H$ and $F$}\label{app:qmoexample}

\begin{figure*}[t]
\centering
\includegraphics[width=0.495\textwidth]{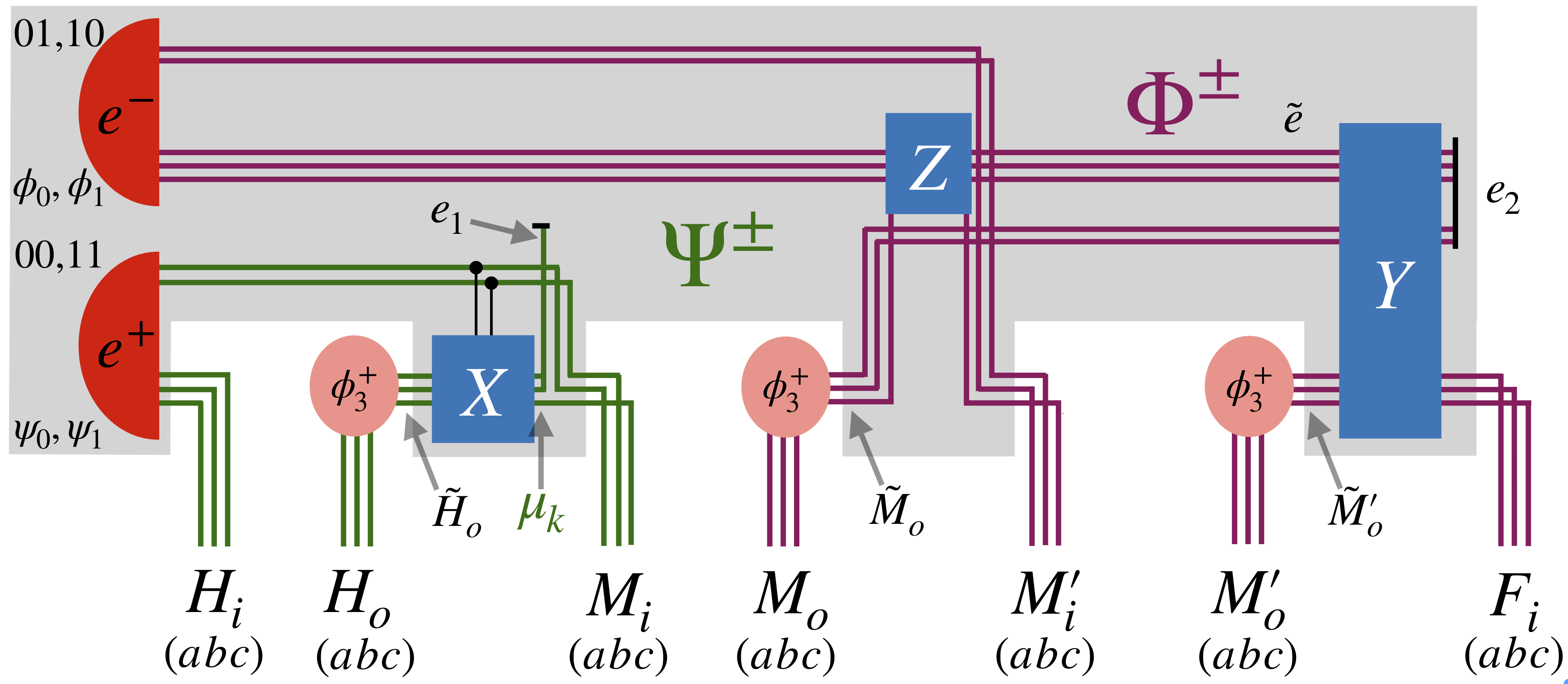}
\includegraphics[width=0.495\textwidth]{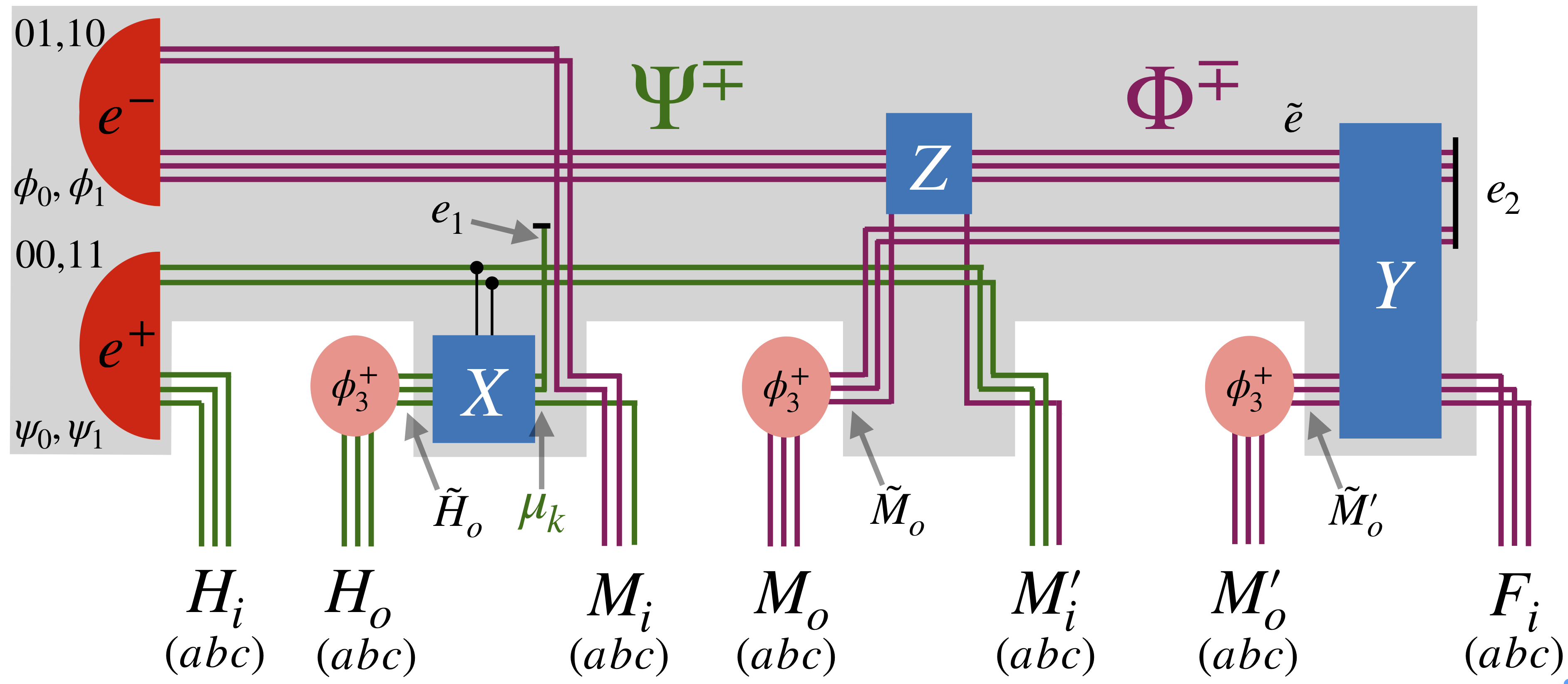}
\caption{\emph{A process with finite quantum Markov order with parts of $M$ kept by $H$ and $F$.} On the left is the first process, described in Eq.~\eqref{eq:processonesm}, in which part of the common cause state $\ket{e^+}$ is sent to $M_{\inp}$ and $\ket{e^-}$ is sent to $M'_{\inp}$. On the right is the second process, described in Eq.~\eqref{eq:processtwosm}, which has the recipients flipped. The process tensor is depicted in gray, and entanglement between parties color-coded in green and mauve. The overall process is a probabilistic mixture of both scenarios. See also Ref.~\cite{Milz2021} for discussion on this process.} \label{fig:qmoexample}
\end{figure*}

Consider the process depicted in Fig.~\ref{fig:qmoexample}, which requires leaving parts of $M$ attached to $H$ and $F$. We label the input/output spaces associated to each time of the process as follows $\{H_{\inp}, H_{\out}, M_{\inp}, M_{\out}, M'_{\inp}, M'_{\out}, F_{\inp}\}$. At each time, the system of interest comprises three qubits, and so each Hilbert space is of the form $\mathcal{H}_X = \mathcal{H}^a_X \otimes \mathcal{H}^b_X \otimes \mathcal{H}^c_X$, where $X$ takes values for the times and $a,b,c$ are labels for the three qubits; whenever we refer to an individual qubit, we will label the system appropriately, e.g., $M_{\inp}^{a}$ refers to the $a$ qubit of the system $M_{\inp}$; whenever no such label is specified, we are referring to all three qubits. See also Ref.~\cite{Milz2021} for discussion on this process.


The environment first prepares the five-qubit common cause states 
\begin{align}
 \ket{e^+} &= \frac{1}{\sqrt{2}}(\alpha \ket{\psi_0,00}+ \beta \ket{\psi_1,11})  
 \quad \mbox{and} \quad \notag \\
 \ket{e^-} &= \frac{1}{\sqrt{2}}(\gamma \ket{\phi_0,01}+ \delta \ket{\phi_1,10}).
\end{align}
Here we have separated the first register, which is a three qubit state, from the second, which consists of two qubits, with a comma. The first parts of the states $\ket{e^+}$ and $\ket{e^-}$ are respectively sent to $H_{\inp}$ and $F_{\inp}$. The second parts are sent either to $M_{\inp}$ or $M'_{\inp}$, according to some probability distribution.

Let the state input at $H_{\out}$ be the first halves of three maximally entangled states $\bigotimes_{x\in \{a,b,c\}} \ket{\phi^+}_{H^x_{\out} \widetilde{H}^x_{\out}}$ with $\ket{\phi^+} := \tfrac{1}{\sqrt{2}}(\ket{00} +\ket{11})$; here, the tilde denotes systems that are fed into the process, whereas the spaces without a tilde refer to systems kept outside of it. The states input at $M_{\out}$, and $M'_{\out}$ are labeled similarly. In between times $H_{\out}$ and $M_{\inp}$, the process makes use of the second part of the common cause state $\ket{e^+}$ to apply a controlled quantum channel $X$, which acts on all three qubits $a,b,c$. Following this, qubits $a$ and $b$ are discarded. The $ab$ qubits input at $M_{\out}$, as well as all three qubits input at $M'_{\out}$ are sent forward into the process, which applies a joint channel $Y$ on all of these systems, as well as the first part of the common cause state $\ket{e^-}$. Three of the output qubits are sent out to $F_{\inp}$, and the rest are discarded. The $c$ qubit input at $M_{\out}$ is sent to $M'_{\inp}$, after being subjected to a channel $Z$, which interacts with the first part of the common cause state $\ket{e^-}$, i.e., the $\phi_0,\phi_1$ register.

Consider the process where $\ket{e^+}$ is sent to $H_{\inp}$ and $M_{\inp}$ and $\ket{e^-}$ to $M'_{\inp}$ and $F_{\inp}$. The process tensor for this case is 
\begin{widetext}
\begin{align}\label{eq:processonesm}
&    \Upsilon^{\pm} = \Psi^{\pm}_{H_{\inp}H_{\out}M_{\inp}} \otimes \Phi^{\pm}_{M_{\out}M'_{\inp}M'_{\out}F_{\inp}},\notag \\
& \mbox{with} \quad
\Psi^{\pm}_{H_{\inp}H_{\out}M_{\inp}} =\mbox{tr}_{e_1}\left[\ket{G^{\pm}}\bra{G^{\pm}} \right] \quad \mbox{and} \quad
\Phi^{\pm}_{M'_{\inp}M'_{\out}F_{\inp}} =\mbox{tr}_{e_2}\left[\ket{K^{\pm}}\bra{K^{\pm}} \right], \, \mbox{where} \notag\\
& \ket{G^{\pm}} = \frac{1}{\sqrt{2}} \left(
\alpha \ket{\psi_0}_{H_{\inp}}
 \ket{\mu_{0}}_{H_{\out} e_1 M_{\inp}^c}\ket{00}_{M_{\inp}^{ab}}
+\beta
 \ket{\psi_1}_{H_{\inp}}
 \ket{\mu_{1}}_{H_{\out} e_1 M_{\inp}^c}\ket{11}_{M_{\inp}^{ab}} \right),
\, \mbox{with} \,\, \notag
\ket{\mu_{k}}_{H_{\out} e_1 M^c_{\inp}} = X^{k}_{\tilde{H}_{\out} \to e_1 M^c_{\inp}} \ket{\phi^+}^{\otimes 3}_{H_{\out} \tilde{H}_{\out}} 
 \\
& \ket{K^{\pm}} = \frac{1}{\sqrt{2}} Y_{\tilde{M}^{ab}_{\out} \tilde{M}'_{\out} \tilde{e} \to F_{\inp} e_2} 
Z^{k}_{\tilde{M}^c_{\out}\tilde{e} \to M'^{c}_{\inp}\tilde{e}}
\left(
\gamma \ket{01}_{M'^{ab}_{\inp}} \ket{\phi_0}_{\tilde{e}} +\delta \ket{10}_{M'^{ab}_{\inp}} \ket{\phi_1}_{\tilde{e}}\right) 
\ket{\phi^+}^{\otimes 3}_{M_{\out} \tilde{M}_{\out}} 
\ket{\phi^+}^{\otimes 3}_{M'_{\out} \tilde{M}'_{\out}}.
\end{align}

Next, consider the process where $\ket{e^+}$ is sent to $H_{\inp}$ and $M'_{\inp}$ and $\ket{e^-}$ to $M_{\inp}$ and $F_{\inp}$. The process tensor for this scenario is 

\begin{align}\label{eq:processtwosm}
&\Upsilon^{\mp} = \Psi^{\mp}_{H_{\inp} H_{\out} M^c_{\inp} M'^{ab}_{\inp}} \otimes \Phi^{\mp}_{M^{ab}_{\inp} M_{\out} M'^c_{\inp} M'_{\out} F_{\inp}}, \notag\\
& \mbox{with} \quad
\Psi^{\mp}_{H_{\inp} H_{\out} M^c_{\inp} M'^{ab}_{\inp}} =\mbox{tr}_{e_1}\left[\ket{G^{\mp}}\bra{G^{\mp}} \right] \quad \mbox{and} \quad
\Phi^{\mp}_{M^{ab}_{\inp} M_{\out} M'^c_{\inp} M'_{\out} F_{\inp}} =\mbox{tr}_{e_2}\left[\ket{K^{\mp}}\bra{K^{\mp}} \right], \, \mbox{where} \notag\\
& \ket{G^{\mp}} = \frac{1}{\sqrt{2}}   \left(
\alpha \ket{\psi_0}_{H_{\inp}}
 \ket{\mu_{0}}_{H_{\out} e_1 M^c_{\inp}}\ket{00}_{M'^{ab}_{\inp}}
+\beta
 \ket{\psi_1}_{H_{\inp}}
 \ket{\mu_{1}}_{H_{\out} e_1 M^c_{\inp} }\ket{11}_{M'^{ab}_{\inp}} \right), 
\, \mbox{with} \,\, \notag
 \ket{\mu_{k}}_{H_{\out} e_1 M^c_{\inp}} = X^{k}_{\tilde{H}_{\out} \to e_1 M^c_{\inp}} \ket{\phi^+}^{\otimes 3}_{H_{\out} \tilde{H}_{\out}}
 \\
& \ket{K^{\mp}} = \frac{1}{\sqrt{2}} Y_{\tilde{M}^{ab}_{\out} \tilde{M}'_{\out} \tilde{e} \to F_{\inp} e_2}  
Z^{k}_{\tilde{M}^c_{\out}\tilde{e} \to M'^{c}_{\inp}\tilde{e}}
\left(
\gamma \ket{01}_{M^{ab}_{\inp}} \ket{\phi_0}_{\tilde{e}}
+\delta \ket{10}_{M^{ab}_{\inp}} \ket{\phi_1}_{\tilde{e}} \right) 
\ket{\phi^+}^{\otimes 3}_{M_{\out} \tilde{M}_{\out}} 
\ket{\phi^+}^{\otimes 3}_{M'_{\out} \tilde{M}'_{\out}}.
\end{align}
\end{widetext}

In the first case, there is entanglement between $H_{\inp \out}$ and $M_{\inp}$, as well as between $M_{\out }M'_{\inp \out}$ and $F_{\inp}$. In the second case, there is entanglement between $H_{\inp \out}$ and $M^c_{\inp}M'^{ab}_{\inp}$, as well as between $M^{ab}_{\inp}M_{\out}M'^c_{\inp} M'_{\out}$ and $F_{\inp}$. The overall process is the average of these two, which will still have entanglement across the same cuts for generic probability distributions that the common cause states are sent out with. 

This process has vanishing memory strength $\Theta$ because we can make a parity measurement on the $ab$ parts of $M_{\inp}$ and $M'_{\inp}$. The parity measurement applies two controlled phases to an ancilla initially prepared in the state $\ket{+}$, with the control registers being qubits $a$ and $b$. If the two control qubits are in states $\ket{00}$ or $\ket{11}$, then $\ket{+} \to \ket{+}$. However, if the control qubits are in states $\ket{01}$ or $\ket{10}$, then $\ket{+} \to \ket{-}$. By measuring the final ancilla, which can be perfectly distinguished since it is in one of two orthogonal states, we can know which process we have in a given run; in either case, there are no $FH$ correlations. Lastly note that this process also has vanishing quantum CMI; this agrees with the analysis in Ref.~\cite{Taranto2019A}, as the instrument that erases the history comprises only orthogonal projectors.

\section{Process Recovery Map}\label{app:recovery}

Let $\mathcal{K}$ be the linear invertible map such that $\mathcal{K}: \{\mathsf{D}^{(x_M)}_M\} \leftrightarrow \{\ket{x_M}\bra{x_M}_m\}$. Such a map always exists since the $\{\mathsf{D}^{(x_M)}_M\}$ are linearly independent and the $\{\ket{x_M}\bra{x_M}_m\}$ are in one-to-one correspondence with $\{\textsf{O}^{(x_M)}_M\}$. Let us denote
\begin{align}
    \underline{\Lambda}^{\mathcal{J}_M}_{FMH} &:= \sum_{x_M}\Upsilon_F^{(x_M)}\otimes \mathsf{D}^{(x_M)}_M\otimes\widetilde{\Upsilon}_{H}^{(x_M)} \\\textup{and}  \quad 
     \Lambda^{\mathcal{J}_M}_{FmH} &:= \sum_{x_M}\Upsilon_F^{(x_M)}\otimes \ket{x_M}\bra{x_M}_m\otimes\widetilde{\Upsilon}_{H}^{(x_M)}
\end{align}
so that $\mathcal{K}(\underline{\Lambda}^{\mathcal{J}_M}_{FMH}) = \Lambda^{\mathcal{J}_M}_{FmH}$ and $\mathcal{K}^{\textup{adj.}} (\Lambda^{\mathcal{J}_M}_{FmH}) = \underline{\Lambda}^{\mathcal{J}_M}_{FMH}$. Here, $\Lambda^{\mathcal{J}_M}_{FmH}$ is a quantum Markov chain state~\cite{Fawzi2015,Sutter2016}. Let $\mathsf{R}_{m \to Fm}^{\mathcal{J}_M}$ be Petz's recovery map such that
\begin{gather}
    \mathsf{R}^{\mathcal{J}_M}_{m \to Fm} (\Lambda^{\mathcal{J}_M}_{mH}) = \Lambda^{\mathcal{J}_M}_{FmH}.
\end{gather}
For this recovery map, the positivity of its argument is important~\cite{Petz2003}. This is why we need to define $\mathcal{K}$, as it turns a possibly non-positive restricted process tensor $\underline{\Lambda}_{FMH}^{\mathcal{J}_M}$ into the positive semidefinite Markov chain state $\Lambda^{\mathcal{J}_M}_{FmH}$, for which the Petz recovery map is guaranteed to exist. We can now simply define the process recovery map as
\begin{gather}
    \mathcal{R}^{\mathcal{J}_M}_{M\to FM} :=\mathcal{K}^{\textup{adj.}} \mathsf{R}^{\mathcal{J}_M}_{m\to Fm} \mathcal{K}
\end{gather}
so that
\begin{gather}
    \mathcal{R}^{\mathcal{J}_M}_{M \to FM} (\underline{\Lambda}_{MH}^{\mathcal{J}_M}) = \underline{\Lambda}_{FMH}^{\mathcal{J}_M}.
\end{gather}

\section{Easily Computable Memory Strength Proxies}\label{app:proxies}

The measured relative entropy computed on the Choi states of process tensors (see Def.~\ref{def:instrumentrelativeentropy}) is a variant of the measured relative entropy (defined in Ref.~\cite{Piani2009}):
\begin{gather}
    S_\mathbb{M}\left(\rho\|\sigma\right) = \sup_{\mathcal{M}\in \mathbb{M}} S\left(\mathcal{M}[\rho]\middle\|\mathcal{M}[\sigma]\right),
\end{gather}
where $\mathbb{M}$ is a set of measurements $\mathcal{M}[\rho] = \sum_{k}\tr{E_k \rho} \ketbra{k}{k}$, with $\{E_k\}$ a POVM satisfying $\sum_k E_k = \mathbbm{1}$.

It is straightforward to show that $S(\rho\|\sigma)\geq S_{\mathbb{M}}(\rho\|\sigma)$, but it is not immediately clear that a similar inequality holds for $S_{\mathbb{J}}$, since the maps $\mathcal{P}_\mathcal{J}$ involved in Def.~\ref{def:instrumentrelativeentropy} are not trace preserving. However, we will now show that a similar inequality holds:
\begin{lem} \label{lem:instrumentrelentbound}
For any pair of process tensors $\Upsilon$, $\Gamma$,
\begin{gather} \label{eq:monotonicity}
    S\left(\Upsilon\middle\|\Gamma\right)\geq \frac{1}{D^\mathtt{i}} S_\mathbb{J}\left(\Upsilon\middle\|\Gamma\right),
\end{gather}
with $D^\mathtt{i} := \prod_{k} d^\mathtt{i}_k$ the total input space dimension of the process tensors.
\end{lem}

\noindent\textit{Proof.} This proof relies on the monotonicity of the relative entropy under the relevant set of CP maps. The crucial property a CP map $\mathcal{E} : \mathcal{B}(\mathcal{H})\rightarrow \mathcal{B}(\mathcal{H}')$ (for a pair of Hilbert spaces $\mathcal{H}$ and $\mathcal{H}'$) must  satisfy in order that $S(\mathcal{E}[\rho]\|\mathcal{E}[\sigma])\leq S(\rho\|\sigma)$ is that both it and its adjoint map $\overline{\mathcal{E}}: \mathcal{B}(\mathcal{H}')\rightarrow \mathcal{B}(\mathcal{H})$ satisfy the Cauchy-Schwarz inequality~\cite{Petz2003}, i.e,
\begin{align}
    \mathcal{E}[X]^\dagger\mathcal{E}[X]\leq& \mathcal{E}(X^\dagger X), \\
    \text{and} \quad \overline{\mathcal{E}}[x]^\dagger\overline{\mathcal{E}}[x]\leq& \overline{\mathcal{E}}(x^\dagger x),
\end{align}
for any bounded operators $X\in\mathcal{B}(\mathcal{H})$ and $x\in\mathcal{B}(\mathcal{H}')$.
This is satisfied when $\|\mathcal{E}[\mathbbm{1}]\|_{\rm op}\leq 1$ and $\|\mathcal{\overline{E}}[\mathbbm{1}]\|_{\rm op}\leq 1$, with $\|X\|_{\rm op}:=\max\{|\lambda|: X- \lambda \mathbbm{1} \;\text{is not invertible}\}$, i.e., the largest singular value of $X$~\cite{Paulsen2002}. If $\mathcal{E}[X] = \sum_k E_k X E_k^\dagger$, then this is equivalent to the following constraints on the Kraus operators: $\sum_{k}E_k E_k^\dagger \leq \mathbbm{1}$ and $\sum_{k}E_k^\dagger E_k \leq \mathbbm{1}$.    

For the maps $\mathcal{P}_\mathcal{J}$ appearing in Def.~\ref{def:instrumentrelativeentropy}, it is possible to show that these conditions do not hold. However, we can write $\mathcal{P}_\mathcal{J} = D^\mathtt{i}  \mathcal{P}_{\hat{\mathcal{J}}}$, where
\begin{gather}
    \mathcal{P}_{\hat{\mathcal{J}}}[X] = \sum_x \tr{\frac{1}{D^\mathtt{i} }\mathsf{O}^{(x)} X}\ketbra{x}{x},
\end{gather}
such that the Kraus operators of $\mathcal{P}_{\hat{\mathcal{J}}}$ take the form $E_{xk} = \ketbra{x}{k}\sqrt{\mathsf{O}^{(x)}/D^\mathtt{i} }$, where $\{\ket{k}\}$ is an orthonormal basis for the space the $\mathsf{O}^{(x)}$'s act on. Since the trace conditions required for the overall action of any instrument imply that $\tr{\mathsf{O}^{\mathcal{J}}} = \sum_x \tr{\mathsf{O}^{(x)}}= D^\mathtt{i}$, we then have $\sum_{xk} E_{xk}E^\dagger_{xk} = \sum_{x} (\tr{\mathsf{O}^{(x)}}/\sum_x \tr{\mathsf{O}^{(x)}}) \ketbra{x}{x}\leq\mathbbm{1}$. For the other condition, we have  $\sum_{xk} E^\dagger_{xk}E_{xk} = \sum_{x}\mathsf{O}^{(x)} / D^\mathtt{i} = \mathsf{O}^{\mathcal{J}}/\tr{\mathsf{O}^{\mathcal{J}}}\leq \mathbbm{1}$ since $\mathsf{O}^{\mathcal{J}}$ is a positive operator and hence the sum of its singular values (its trace) cannot be smaller than its largest. Therefore, $\mathcal{P}_{\hat{\mathcal{J}}}$ satisfies the Cauchy-Schwarz inequality in both directions and we have
\begin{gather} \label{eq:Dmonotonic}
    S\left(\mathcal{P}_{\hat{\mathcal{J}}}[\Upsilon]\middle\|\mathcal{P}_{\hat{\mathcal{J}}}[\Gamma]\right) = \frac{1}{D^\mathtt{i}}S\left(\mathcal{P}_{{\mathcal{J}}}[\Upsilon]\middle\|\mathcal{P}_{{\mathcal{J}}}[\Gamma]\right) \leq S\left(\Upsilon\middle\|\Gamma\right),
\end{gather}
where, in the first equality, we have used the simply demonstrated fact that $S(\alpha \rho\|\alpha \sigma) = \alpha S( \rho\| \sigma)$ for any scalar $\alpha$. Since Eq.~\eqref{eq:Dmonotonic} holds for any $\mathcal{J}\in\mathbb{J}$, Eq.~\eqref{eq:monotonicity} follows directly from Def.~\ref{def:instrumentrelativeentropy}. \hfill $\square$

Since the measured relative entropy on the r.h.s. of Eq.~\eqref{eq:monotonicity} is equal to the memory strength (see Prop.~\ref{prop:loosethetabound}), one can replace the r.h.s. of Thms.~\ref{thm:recovery} and~\ref{thm:ic} and Cor.~\ref{cor:densityop} by the proxy memory strength $D_{FH} S(\Upsilon_{FmH}^{\mathcal{J}_M}\|\underline{\Lambda}^{\mathcal{J}_M}_{FmH})$, where $\Upsilon_{FmH}^{\mathcal{J}_M} = \sum_{x_M} \widetilde{\Upsilon}_{FH} \otimes \ket{x_M}\bra{x_M}_m$ is the projected process tensor onto the classical pointer basis $m$ and similarly for $\underline{\Lambda}^{\mathcal{J}_M}_{FmH}$, and $D_{FH}$ is the dimension that the future-history process tensor acts upon. This is precisely what is plotted as a proxy for memory strength in Fig.~\ref{fig:memstrengthplot} of the main text. The advantage of doing so is that the proxy does not require computing an optimization over future and history instruments, therefore making it easier to calculate. The disadvantage is that the bound is looser than the bound based upon the optimized memory strength by a factor of $D_{FH}$. Importantly, this is smaller than the full system-environment dimension as it only accounts for the environmental impact on the system dynamics and can be tomographically reconstructed. Note further that the complexity of $FH$ is often a choice: for a given $d$-dimensional system, one may be interested in only predicting its instantaneous state at some future time following an initial (historic) preparation for some memory instruments---in this case, $D_{FH} = d^2 \times d^2$ (in general, the dimension of a process tensor grows exponentially in the number of timesteps). If the memory strength vanishes, the recovered process can be used to perfectly simulate the dynamics arbitrarily far into the future and for any number of timesteps.

Alternatively, by restricting the measured relative entropy to the the special class of unbiased instruments $\mathcal{J}^{{\rm ub}}\in \mathbb{J}^{{\rm ub}}$ (on $FH$) with deterministic action satisfying $\mathsf{O}^{\mathcal{J}^{{\rm ub}}} = \mathbbm{1}/D^\mathtt{o}$, we can prove the following bound:
\begin{prop} \label{prop:thetabound}
For any $\Upsilon_{FMH}$, $\mathcal{J}_M$ and $\underline{\Lambda}^{\mathcal{J}_M}_{FMH}$ as defined in Eq.~\eqref{eq:restricted} of the main text,
\begin{gather} \label{eq:thetabound}
    S(\Upsilon_{FmH}^{\mathcal{J}_M} \| \underline{\Lambda}_{FmH}^{\mathcal{J}_M}) \geq S_{{\mathbb{J}^{{\rm ub}| \mathcal{J}_M}}}\left(\Upsilon_{FMH}\middle\|\underline{\Lambda}^{\mathcal{J}_M}_{FMH}\right),
\end{gather}
with $\mathbb{J}^{{\rm ub}| \mathcal{J}_M} = \mathbb{J}\cap {\rm span}(\mathcal{J}_M)\cap \mathbb{J}_{FH}^{\rm ub}$ the set of instruments whose elements have support on $M$ only in the linear span of the elements of $\mathcal{J}_M$ and for which the $FH$ part is unbiased.
\end{prop}
\noindent \textit{Proof.} First we note that, for unbiased instruments,
\begin{align}\notag
    \mathcal{P}_{\mathcal{J}^{\rm ub}}[\Upsilon] =& \sum_{x}\tr{\mathsf{O}^{(x)} \Upsilon}\ketbra{x}{x} \\
    =& \sum_{x}\tr{D^{\mathtt{o}}\mathsf{O}^{(x)} \hat{\Upsilon}}\ketbra{x}{x},
\end{align}
where the trace-normalized Choi state $\hat{\Upsilon}$ is a physical density operator and $\sum_x D^{\mathtt{o}}\mathsf{O}^{(x)} = \mathbbm{1}$, i.e., the rescaled instrument elements form a POVM. In this case, the action of the instrument map $\mathcal{P}_{\mathcal{J}^{\rm ub}}$ on the process tensor looks like a trace-preserving measurement map on the normalized Choi state, for which the usual monotonicity of relative entropy holds. That is, for process tensors $\Upsilon$ and $\Gamma$, 
\begin{gather}
    S\left(\hat{\Upsilon}\middle\|\hat{\Gamma}\right)\geq S\left(\mathcal{P}_{\mathcal{J}^{\rm ub}}[\Upsilon]\middle\|\mathcal{P}_{\mathcal{J}^{\rm ub}}[\Gamma]\right).
\end{gather}

We can therefore follow the same steps as in the proof of Prop.~\ref{prop:loosethetabound} to arrive at
\begin{gather}
    S(\Upsilon_{FmH}^{\mathcal{J}_M} \| \underline{\Lambda}_{FmH}^{\mathcal{J}_M})
    \geq S_{\mathbbm{J}_{FmH}^{\rm ub}}\left({\Upsilon}^{\mathcal{J}_M}_{FmH}\middle\|{\Lambda}_{FmH}^{\mathcal{J}_M}\right),
\end{gather}
where $\mathbbm{J}_{FmH}^{\rm ub}$ is the set of $FmH$ instruments $\mathcal{J}_{FmH}^{\rm ub}$ satisfying $\mathsf{O}^{\mathcal{J}_{FmH}^{\rm ub}} = \mathbbm{1}_{FmH} /D^\mathtt{o}_{FH}$. In the same manner as the full set of instruments on $FmH$ can be identified with the set of instruments $\mathbb{J}\cap {\rm span}(\mathcal{J}_M)$, $\mathbbm{J}_{FmH}^{\rm ub}$ can be identified with $\mathbb{J}\cap {\rm span}(\mathcal{J}_M)\cap \mathbb{J}_{FH}^{\rm ub}$, and Eq.~\eqref{eq:thetabound} follows accordingly. \hfill $\square$

\section{Details for Case Study}\label{app:casestudy}

The authors of Ref.~\cite{Pang2017} have examined the non-Markovianity of the system dynamics for the model described in the main text using the breakdown of CP-divisibility and the increase of the trace-distance distinguishability between arbitrary input states (note that the latter is implied by the former~\cite{Breuer2016}) as measures of the existence of temporal correlations. In this study, it was found that there is an abrupt transition between CP-divisible and non-CP-divisible dynamics in the parameter regime. However, such two-time considerations are insufficient to characterize the full memory effects of the process, as they necessarily fail to capture multi-time effects. Before we go on to examine the memory length of said process, we first show that it is non-Markovian for all parameters in the considered regime using a recently introduced notion of non-Markovianity that accounts for all (multi-time) temporal correlations~\cite{Pollock2018L}. See also Ref.~\cite{TarantoThesis} for further details on this case study.

\begin{figure*}[t]
\centering
\includegraphics[width=\textwidth]{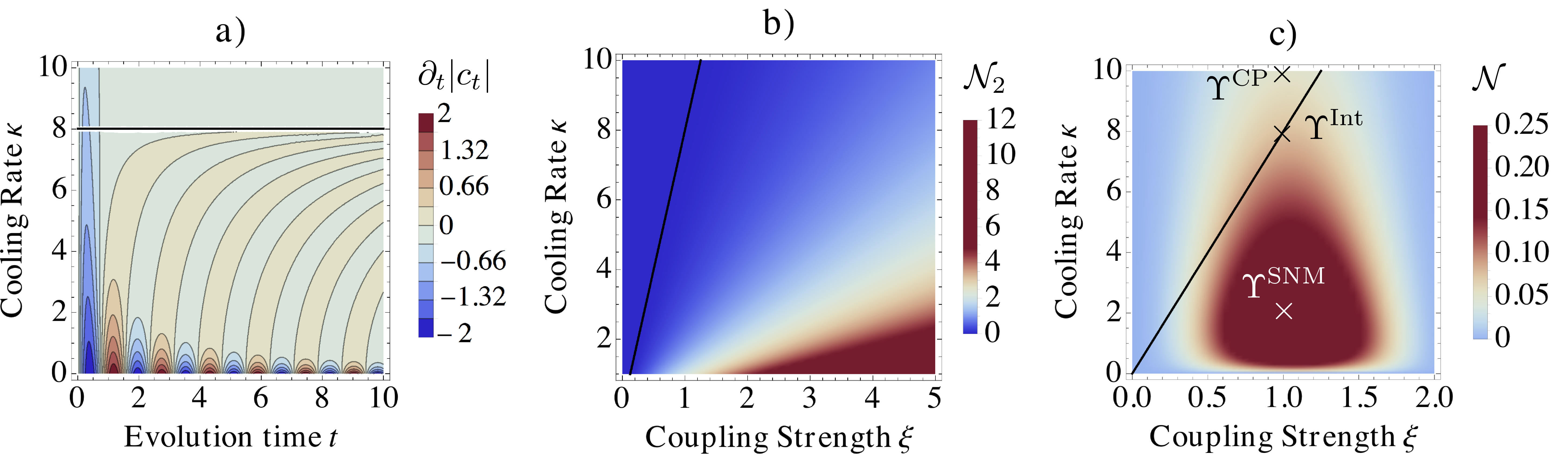}
\caption{\emph{Abrupt transition between CP-divisible and non-divisible dynamics}~\cite{TarantoThesis}. In panel \textbf{a)}, we plot $\partial_t |c_t| $ with $\xi = 1$. As $\sgn{(\tfrac{\dot{c}_t}{c_t})} = \sgn{(\partial_t |c_t|)}$, this implies the dynamics is CP-divisible for $\kappa \geq 8$, but not for $\kappa < 8$. In particular, there is an abrupt transition along the line $\kappa = 8$. In panel \textbf{b)}, we plot the two-time non-Markovianity $\mathcal{N}_{2}$ as per Eq.~\eqref{eq:nmtwotime}. This is plotted in the parameter space $\xi \in [0,5]$ and $ \kappa \in [0,10] $. Note that this measure of non-Markovianity vanishes for everything above the black line $\kappa = 8 \xi$. In panel \textbf{c)} we plot the multi-time non-Markovianity $\mathcal{N}$ of $\Upsilon_{6:1}(\xi,\kappa)$. An important distinction to note is that the two-time non-Markovianity in panel \textbf{b)} results from an integration of positive memory contributions over all times, whereas the multi-time non-Markovianity in panel \textbf{c)} is computed for fixed process tensors in the parameter space. Although the non-Markovianity is small above the line $\kappa = 8 \xi$, it is non-zero. Moreover, there is no abrupt transition between regimes, as all memory effects are accounted for. We also depict with crosses the three specific process tensors defined in Eq.~\eqref{eq:processregimes} for which we calculate the instrument-specific memory strength in the main text.} \label{fig:cpdiv}
\end{figure*}

Due to the simplicity of the model, the analytic form of the equation of motion on the level of the system alone can be derived and is written as~\cite{Pang2017} 
\begin{align}
	\frac{\partial \rho^{S}_t}{\partial t} = - \frac{\dot{c}_t}{2 c_t} \mathcal{L}[\sigma_{x}^S]( \rho^{S}_t),
\end{align}
where the time-dependent coefficient is given by
\begin{widetext}
\begin{align}
c_t =  \begin{cases}
	 \exp{\left(-\frac{\kappa t}{4} \right)}\left( \frac{\kappa \sinh{\frac{t}{4} \sqrt{\kappa^2 - 64 \xi^2} } }{\sqrt{\kappa^2 - 64 \xi^2}} + \cosh{\frac{t}{4} \sqrt{\kappa^2 - 64 \xi^2}} \right) &\text{for } \kappa^2 > 64 \xi^2 \\
	 \exp{\left(-\frac{\kappa t}{4} \right)}\left( \frac{\kappa \sin{\frac{t}{4} \sqrt{64 \xi^2- \kappa^2} } }{\sqrt{64 \xi^2- \kappa^2}} + \cos{\frac{t}{4} \sqrt{64 \xi^2- \kappa^2}} \right) &\text{for } \kappa^2 < 64 \xi^2 \\
	\exp{\left(-\frac{\kappa t}{4} \right)} \left( 1 + \frac{1}{4} \kappa t\right) &\text{for } \kappa^2 = 64 \xi^2.
\end{cases}
\end{align}
\end{widetext}
A necessary and sufficient criterion for the dynamics to be CP-divisible is that the coefficients of the dissipation terms in the above master equation for the system, i.e, $- \frac{\dot{c}_t}{2 c_t}$, are non-negative for all times~\cite{Breuer2016}. Explicit calculation shows that for $\kappa^2 \geq 64 \xi^2$, $- \frac{\dot{c}_t}{2 c_t}$ is always non-negative, whereas for $\kappa^2 < 64 \xi^2$, $- \frac{\dot{c}_t}{2 c_t}$ is negative whenever $\cot{\tfrac{1}{4} t \sqrt{64 \xi^2 - \kappa^2}} < - \tfrac{\kappa}{\sqrt{64 \xi^2 - \kappa^2}}$. We therefore see an abrupt transition between CP-divisible and non-CP-divisible dynamics across the line $\kappa^2 = 64 \xi^2$, as shown here in Fig.~\ref{fig:cpdiv}a). 


In the CP-divisible regime, the trace-distance between any two states subject to the evolution is always non-increasing~\cite{Breuer2016}. This fact allows for the total quantification of two-time non-Markovianity $\mathcal{N}_2$ by integrating any increases in the trace-distance over all time, which has been shown in Ref.~\cite{Pang2017} to yield the analytic result
\begin{align}\label{eq:nmtwotime}
	\mathcal{N}_2 = \frac{1}{\exp{\left( \frac{\kappa \pi}{ \sqrt{64 \xi^2 - \kappa^2} } \right) } -1},
\end{align}
for $\kappa^2 < 64 \xi^2$ and zero otherwise. This is plotted in Fig.~\ref{fig:cpdiv}b).

However, CP-divisibility does not imply Markovianity~\cite{Milz2019}; as such, Figs.~\ref{fig:cpdiv}a) and b) do not provide a comprehensive picture of the many prevalent memory effects for different choices of parameters $\kappa$ and $\xi$. Here, we explicitly calculate the process tensor for the dynamics and show that it is non-Markovian for the entire parameter regime, before exploring the behavior of the instrument-specific memory strength introduced in the main text. 

We consider a parameter grid $\xi \in [0,2]$ and $ \kappa \in [0,10]$ with increments of $0.1$ in each direction and construct the $n=6$ step process tensor, $\Upsilon_{6:1}(\xi,\kappa)$. Here, for simplicity, we assume an initially uncorrelated system-environment state, such that the process tensor begins on an output space. We also choose uniform spacing between timesteps of $dt = 0.3$, which corresponds to the natural timescale over which the trace-distance between arbitrary initial system states increases for most values in the parameter space~\cite{Pang2017}. This means that the final time of the process tensor is $T=1.5$, which corresponds to where the CP-divisibility criteria would witness non-Markovianity for a wide range of parameters.

At each point, we calculate the multi-time non-Markovianity $\mathcal{N}$ in the process by considering the distance between the process tensor and its nearest Markovian counterpart~\cite{Pollock2018L}. Here, we choose the pseudo-distance to be the relative entropy, $\mathcal{N}:=\mathcal{D}(\Upsilon_{6:1}(\xi,\kappa) \| \Upsilon^\textup{Markov}_{6:1})$, in which case the minimum occurs for the Markovian process that is simply built up from the marginals of the original process tensor~\cite{Vedral2002}, i.e., using the relative entropy circumvents the normally necessary minimization. The corresponding results are depicted in Fig.~\ref{fig:cpdiv}c), which indicates that the process is non-Markovian for all parameters in the chosen range. In particular, there is no abrupt transition between regimes. Although the non-Markovianity is small above the line $\kappa = 8 \xi$---across which the dynamics transitions from CP-divisible to non-CP-divisible---it is non-zero, indicating a weak but present memory. The two-time witness of non-Markovianity in Eq.~\eqref{eq:nmtwotime} is insensitive to such effects, which leads to the abrupt transition between regimes; by capturing all multi-time correlations, the non-Markovianity calculated via the process tensor shows this transition to be artificial. 
This begs the question: how long does the memory persist? In the main text, we study the behavior of memory with respect to a number of instruments of interest~\cite{TarantoThesis}, namely: 

i) The `do nothing' or identity instrument, which intuitively corresponds to the natural memory strength of the process as the system is not actively intervened on throughout its evolution. This single `outcome' instrument sequence on $M$ consists of $\ell$ identity maps, whose Choi state is $\Psi^{+ }_M = \bigotimes_{j=k-\ell}^{k-1} \Psi^{+ }_{j}$, the natural memory strength is quantified by the correlations in $\Upsilon_{FH}^{\mathcal{I}_M} := \ptr{M}{\Psi^{+ }_M \Upsilon_{FMH}}$ (the memory strength with respect to any unitary transformations can be defined similarly). 

ii) The causal break instrument comprising of an informationally complete set of independently measured and reprepared states. The causal break sequence chosen is a symmetric single-qubit informationally-complete POVM comprising elements
\begin{align}\label{eq:povm}
    \Pi^{(x)} := \tfrac{1}{4}(\mathbbm{1}+\tfrac{1}{\sqrt{3}} \textstyle{\sum_i}\alpha_i^{(x)} \sigma_i),
\end{align}
where $\{ \alpha^{(x)} \} = \{ (1,1,1),(1,-1,-1),(-1,1,-1),\allowbreak(-1,-1,1)\}$ are tetrahedral coordinate vectors, followed by the independent repreparation into one of the set of states (with uniform probability) $\{ \ket{0}\bra{0}, \ket{1}\bra{1}, \ket{+}^x\bra{+}^x, \ket{+}^y\bra{+}^y\}$, where $\ket{+}^{x/y}$ is the $+1$ eigenstate of $\sigma_x / \sigma_y$. Aggregating the correlations for each outcome to the corresponding instrument level provides another notion of memory strength amenable to our framework; we take the average with respect to the probability distribution $p_x = \tr{\mathsf{B}_M^{(x_m) \textup{T}} \Upsilon_{FMH}}$, where $\mathsf{B}_M^{(x_m)}$ are the elements of the causal break instrument above.

iii) The completely-noisy instrument, which discards the output and reprepares the maximally-mixed state, capturing the strength of noise-resistant memory. The Choi state of this (single-outcome) instrument is the identity matrix and so the correlations in the marginal process tensor $\Upsilon_{FH}^{\mathbbm{1}_M} = \Upsilon_{FH} = \tfrac{1}{D^{\out}_M} \ptr{M}{\Upsilon_{FMH}}$ quantify this type of memory. 

We focus on the instrument-specific memory strength for the instruments above for three specific process tensors, one in each regime of interest for the dynamics described, chosen as [see Fig.~\ref{fig:cpdiv}c)]
\begin{align}\label{eq:processregimes}
	 \Upsilon^{\textup{CP}} :=& \Upsilon_{6:1}(1,10), \notag \\
	 \Upsilon^{\textup{Int}} :=& \Upsilon_{6:1}(1,8), \notag \\
	 \Upsilon^{\textup{SNM}} :=& \Upsilon_{6:1}(1,1).
\end{align}
In Fig.~\ref{fig:memstrengthplot} of the main text, we consider the three process tensors above upon which the identity, completely noisy and causal break instruments described previously are applied. We plot the memory strength proxy $\sqrt{2 d^u  S(\Upsilon_{FmH}^{\mathcal{J}_M}\|\underline{\Lambda}^{\mathcal{J}_M}_{FmH})}$ and the difference between expectation values $\left|\langle C\rangle_{\Upsilon_{F\!M\!H}} - \langle C\rangle_{ \underline{\Lambda}^{\mathcal{J}_M}_{F\!M\!H}}\right|$ (i.e., the l.h.s. of Thm.~\ref{thm:recovery}), where $d = 2$ and $u = 4-\ell$ and we choose $C$ to an observable corresponding to an initial preparation of state $\ket{0}$, doing nothing to the process for the middle four timesteps and then finally making a measurement yielding outcome 1 of the POVM defined in Eq.~\eqref{eq:povm}.

\end{document}